\documentclass[aps,pra,onecolumn,final,floatfix,superscriptaddress,nofootinbib,10pt,notitlepage]{revtex4-1}
\usepackage{graphics}
\usepackage{latexsym}
\usepackage{amsmath}
\usepackage{amsfonts}
\usepackage{amssymb}
\usepackage{braket}
\usepackage{graphicx}
\usepackage{dsfont}
\usepackage[bb=boondox]{mathalfa}
\usepackage{simplewick}
\usepackage[normalem]{ulem}
\newcommand{\norm}[1]{\left\lVert#1\right\rVert}

\newcount\Comments  % 0 suppresses notes to selves in text
\usepackage{color}
\definecolor{darkgreen}{rgb}{0,0.5,0}
\definecolor{purple}{rgb}{1,0,1}
\newcommand\redsout{\bgroup\markoverwith{\textcolor{red}{\rule[0.5ex]{2pt}{0.4pt}}}\ULon}
\newcommand{\kibitz}[2]{\ifnum\Comments=1\textcolor{#1}{#2}\fi}
\begin{document}

\title[Title]{Algorithm for initializing a generalized fermionic Gaussian state on a quantum computer}

%-------------------------------------------------------
% author list
\author{Michael P. Kaicher}
\email[Correspondence to: ]{mkaicher(at)ucm.es}
\affiliation{Departamento de F\'isica Te\'orica, Universidad Complutense, 28040 Madrid, Spain}
\affiliation{Theoretical Physics Department, Saarland University, 66123 Saarbr{\"u}cken, Germany}

\author{Simon B. J\"{a}ger}
\affiliation{Theoretical Physics Department, Saarland University, 66123 Saarbr{\"u}cken, Germany}
\affiliation{JILA and Department of Physics, University of Colorado, Boulder, Colorado 80309-0440, USA}

\author{Frank K. Wilhelm}
\affiliation{Theoretical Physics Department, Saarland University, 66123 Saarbr{\"u}cken, Germany}
\affiliation{Institute for quantum computing analytics, Forschungszentrum J\"ulich, 52428 J\"ulich, Germany}
%-------------------------------------------------------
\begin{abstract}
We present explicit expressions for the central piece of a variational method developed by Shi \textit{et al.} (2018 \textit{Ann. Phys.} 390, 245) which extends variational wave functions that are efficiently computable on classical computers beyond mean-field to generalized Gaussian states. In particular, we derive iterative analytical expressions for the evaluation of expectation values of products of fermionic creation and annihilation operators in a Grassmann variable-free representation. Using this result we find a closed expression for the energy functional and its gradient of a general fermionic quantum many-body Hamiltonian. We present a simple gradient-descent-based algorithm that can be used as an optimization subroutine in combination with imaginary time evolution, which by construction guarantees a monotonic decrease of the energy in each iteration step. Due to the simplicity of the quantum circuit implementing the variational state Ansatz, the results of the algorithms discussed here and in Shi \textit{et al.}'s work could serve as an improved, beyond mean-field initial state in quantum computation.
\end{abstract}
%-------------------------------------------------------
\maketitle
%--------------------------------------------------------------------
%--------------------------------------------------------------------
%--------------------------------------------------------------------
%--------------------------------------------------------------------
\section{Introduction\label{intro}}
Quantum computers are believed to be able to solve certain computational problems faster than any known algorithm on a classical computer \cite{manin1980computable,feynman1982simulating}. In particular, one considers the simulation of the low-energy physics of fermionic many-body problems a suitable target where quantum computers could enable the simulation of system sizes well beyond the capabilities of any classical machine, and thus potentially lead to an understanding of a plethora of quantum mechanical phenomena \cite{Lloyd1073}. The computational advantage of a quantum computer stems from its ability to represent a state using a number of qubits that scales as the logarithm of the size of the problem, combined with its capability to execute quantum algorithms that possess a provable speedup over any known classical algorithm \cite{bernstein1997,grover1996fast,simon1997power,shor1999polynomial,berry2015simulating,berry2015hamiltonian,Low2019hamiltonian,low2017optimal}. While quantum computers are well-suited for studying the time evolution of a quantum state, the problem of initializing the quantum computer in a suitable eigenstate of the Hamiltonian---for instance in its ground state---is believed to be exponentially hard in most instances even for quantum computers \cite{ge2019faster}. However, quantum algorithms have been developed that can efficiently approximate the ground state within desired precision \cite{kitaev1995quantum,abrams1999quantum,poulin2009preparing,ge2019faster}. These quantum algorithms still require knowledge about the ground state energy and the ability to  find an initial state that has sufficient support on the true ground state. Even for algorithms which can be executed on current quantum hardware, such as the variational quantum eigensolver, heuristics suggest that preparing the quantum computer in an initial state with large support on the ground state is beneficial \cite{peruzzo2014variational,mcclean2016theory}. Therefore pre-error-corrected, as well as more advanced post-error-corrected quantum algorithms rely on classical methods that allow one to compute an initial state on a classical computer and then implement it fast and reliably on a quantum computer. 

One of the most successful classical methods to study many-body systems is generalized Hartree-Fock theory \cite{Bach1993GeneralizedHT,kraus2010generalized}, which allows one to compute a single Slater determinant whose energy is a local minimum within a mean-field description of the system Hamiltonian. Single Slater determinants belong to the family of fermionic Gaussian states, which are of paramount importance for fermionic systems, as they allow for an efficient computation of expectation values through Wick's theorem \cite{wick1950evaluation,bravyi2005lagrangian} and have already been implemented on superconducting quantum hardware, where they may be employed as an initial state for quantum algorithms \cite{arute2020hartree}. 

While fermionic Gaussian states are able to capture some correlations, they are in general insufficient to accurately describe strongly correlated quantum systems. This can be traced back to the fact that they typically only describe the ground state of non-interacting systems, i.e. quadratic Hamiltonians \cite{kraus2009quantum,hackl2018aspects}. Recently, Shi \textit{et al.} demonstrated that certain families of non-Gaussian fermionic and bosonic states can be computed efficiently using a variational wave function that includes non-factorizable correlations \cite{SHI2018245}, effectively extending the family of efficiently computable states beyond Gaussian states to so-called \textit{generalized Gaussian states}. We will refer to the method employed by Shi \textit{et al.} as the \textit{variational method}. Since then, a number of works have used this variational method in order to classically simulate various spin-, bosonic-, and fermionic systems \cite{sala2018variational,ashida2018solving,ashida2018variational,ashida2019efficient,ashida2019quantum,guaita2019gaussian,shi2019trapped,liu2020non,wang2020theory,wang2020zero,pan2021quantum}. However, to the best of our knowledge, no work has so far applied the variational method of Shi \textit{et al.} to optimize the non-Gaussian part for strongly correlated purely fermionic systems. This can in large part be attributed to the fact that the treatment of non-Gaussian variational parameters for fermionic systems is highly non-trivial both from a mathematical, and a numerical point of view.

\begin{figure}[h!]
    \centering
    \includegraphics[width=0.4\textwidth]{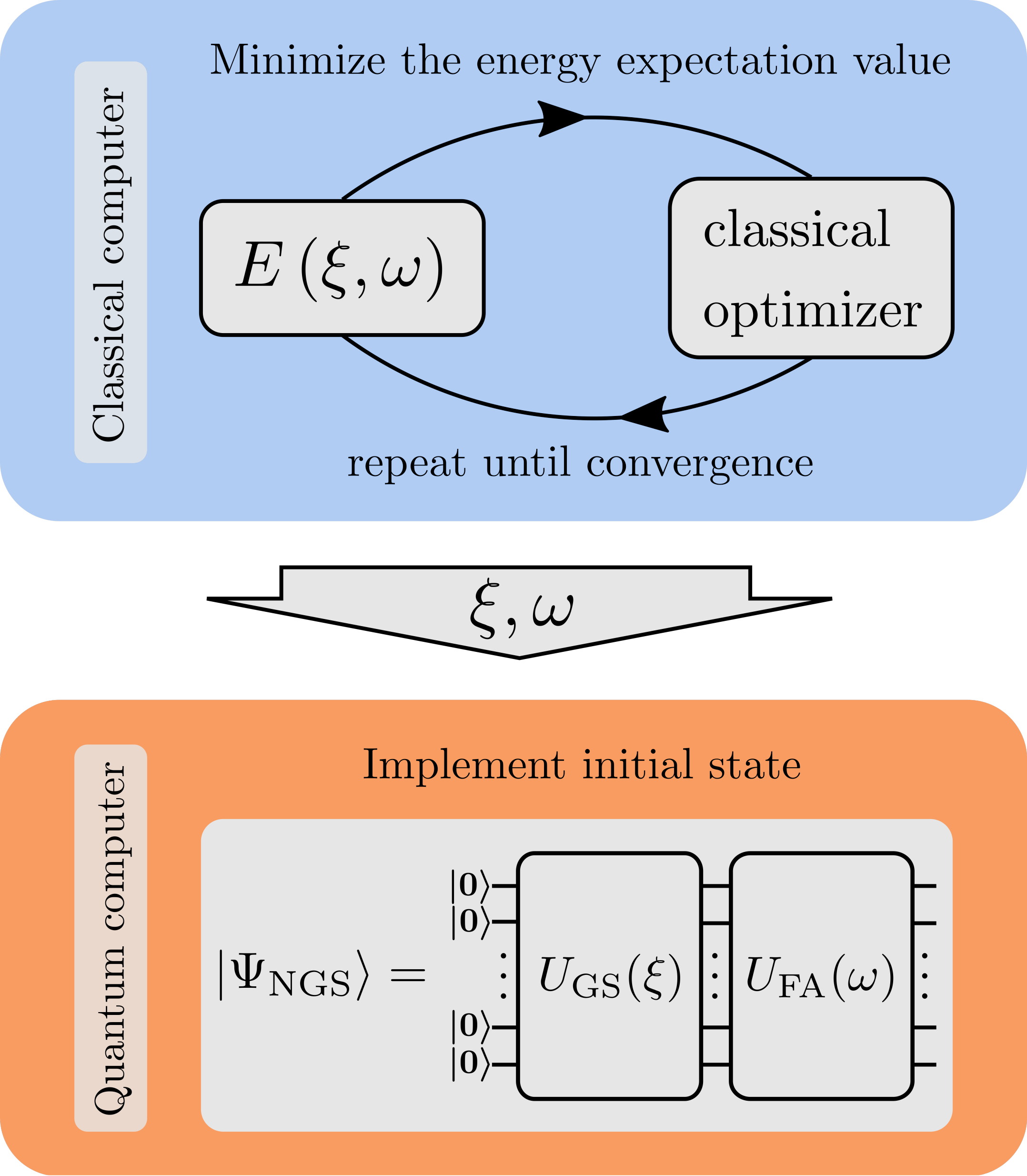}
    \caption{The upper blue box depicts the working principle of the variational method studied here, where a set of parameters $\{\xi,\omega\}$ is optimized in order to find a minimum of the energy functional $E(\xi,\omega)$ as defined in Eq.~\eqref{p350}. The result of the optimization algorithm can be used to initialize a quantum computer in a non-Gaussian state Ansatz, as shown in the orange box, that could in principle improve over generalized Hartree-Fock methods as it includes non-factorizable correlations \cite{SHI2018245}. In contrast to the variational quantum eigensolver method \cite{peruzzo2014variational,mcclean2016theory}, both the optimization algorithm and the computation of the expectation values are performed on a classical computer.}
    \label{fig:var_method}
\end{figure}

Following  \cite{SHI2018245}, we will consider a variational state that is described by two sets of variational parameters $\xi$ and $\omega$, 
\begin{align}
\left|\Psi_{\text{NGS}}(\xi,\omega)\right\rangle= \hat U_{\text{FA}}(\omega)\hat U_{\text{GS}}(\xi)\ket{0},\label{p2}
\end{align}
where $\hat U_{\text{GS}}(\xi)$ is a unitary operator generating a fermionic Gaussian state $\ket{\Psi_{\text{GS}}}=\hat U_{\text{GS}}\ket{0}$, $\hat U_{\text{FA}}(\omega)$ is a unitary operator generating non-factorizable correlations, and $\ket{0}$ denotes a fermionic vacuum state. More precisely,
\begin{align}
\hat U_{\text{GS}}(\xi) 
=&e^{\frac{i}{4}\sum_{j,k=1}^{2N_f}\hat A_j\xi_{jk}\hat A_k}     \label{p6},\\
\hat U_{\text{FA}}(\omega) 
=& e^{\frac{i}{2}\sum_{j, k=1}^{N_f}\omega_{jk}:\hat n_j\hat n_k:},\label{p9}
\end{align}
where $\xi_{jk}$ ($\omega_{jk}$) are the variational parameters collected in the anti-symmetric and Hermitian (symmetric) matrix $\xi$ ($\omega$). Since the diagonal part of $\omega$ describes single particle terms already contained in the Gaussian part, we set $\omega_{jj}=0$ for $j\in\{1,\dots,N_f\}$. Here, $:\hat n_j\hat n_k:$ is the normal-ordered product of the fermionic number operators $\hat n_j=\hat c_j^\dag \hat c_j$, and 
\begin{align}
\hat A_j=&\hat c_j^\dag +\hat c_j, \quad \text{if} \ j\in \{1,\dots,N_f\},\label{ma1}    \\
\hat A_j=&i(\hat c_j^\dag -\hat c_j),\quad \text{if} \ j\in \{N_f+1,\dots,2N_f\},\label{ma2}
\end{align}
are Majorana operators, where $N_f$ denotes the number of fermionic modes. Throughout this work, $\hat c_j^\dag$ and $\hat c_j$ denote  fermionic creation and annihilation operators satisfying the canonical anticommutation relations $\{\hat c_j,\hat c_k\}=0$ and $\{\hat c_j,\hat c_k^\dag\}=\delta_{jk}$, where $\delta_{jk}$ is the Kronecker-delta. 

Central to the variational method is the ability to turn the energy expectation value of the system Hamiltonian $\hat H$,
\begin{align}
E(\xi,\omega)=\braket{\Psi_{\text{NGS}}(\xi,\omega)|\hat H|\Psi_{\text{NGS}}(\xi,\omega)},\label{p350}
\end{align}
into an energy functional that can be evaluated efficiently on a classical computer. The working principle of the variational method is that, provided one is able to evaluate the functional in Eq.~\eqref{p350}, one can use an optimization method to find the set of variational parameters $\{\xi,\omega\}$ that approximate the ground state $\ket{\Psi_0}$ of $\hat H$ within the family of states defined by the variational Ansatz in  Eq.~\eqref{p2}. This is summarized in the upper blue box in Fig.~\ref{fig:var_method}. Once the optimization algorithm has converged, one can use the final set of variational parameters to either study the properties of the state (e.g. perform a real time evolution, or compute observables) on classical computers, or use the result to implement an initial state for a quantum algorithm, as shown in the lower orange box in Fig.~\ref{fig:var_method} and detailed in Section~\ref{qc}.

The main goal of our work is threefold. First, we derive explicit expressions for applying the variational method to strongly interacting, purely fermionic systems. Due to the presence of the non-Gaussian unitary $\hat U_{\text{FA}}$, it is necessary to derive a generalized Wick theorem\footnote{This should not be confused with the computation of the overlap $\braket{\Psi_{\text{GS}}(\xi)|\text{Poly}(\hat C)|\Psi_{\text{GS}}(\xi')}$, where $\text{Poly}(\hat C)$ is a polynomial of fermionic operators, of two possibly distinct fermionic Gaussian states, which is sometimes also called ``generalized Wick theorem''  \cite{balian1969nonunitary}.}, for which we present a Grassmann variable-free representation, thus simplifying the result presented in  \cite{SHI2018245} and adding two expressions required for the computation. Second, while Shi \textit{et al.} choose the equations of motions of both types of variational parameters $\xi$ and $\omega$ following an imaginary time evolution, alternative methods for the optimization of the variables can be used. We discuss how one can use a hybrid approach which lets the Gaussian parameters $\xi$ evolve following an imaginary time evolution, while updating the non-Gaussian parameters $\omega$ through a different gradient-descent-based Ansatz that originates from the principle of ensuring a monotonic decrease of the energy functional in each time step.  We derive closed formulas for the time evolution of $\xi$ (more precisely, for the covariance matrix $\Gamma_m$ that describes the fermionic Gaussian state which is characterized by $\xi$) and $\omega$. At last, we present a $\mathcal O(\log(N_f))$-depth quantum circuit that shows how the classically computed approximate ground state can be implemented on a quantum computer to serve e.g. as an initial state with a potentially improved overlap in comparison to an Hartree-Fock Ansatz, where the latter can be described by a single fermionic Gaussian state $\ket{\Psi_{\text{GS}}}$.

This paper is structured as follows. In Section~\ref{expectation_values}, we present an explicit formula for evaluating a particular type of expectation value for products of fermionic creation and annihilation operators with respect to a non-Gaussian state Ansatz such as Eq.~\eqref{p2}, which is termed \textit{generalized Wick theorem}.  We derive analytical expressions for the energy expectation value of a quantum many-body Hamiltonian with respect to the non-Gaussian state Ansatz  in Section~\ref{qmb}. In Section~\ref{hybrid}, we introduce a simple alternative to the imaginary time evolution-based Ansatz of Shi \textit{et al.}, which only requires the evaluation of the gradient of the energy functional with respect to the non-Gaussian variational parameters and the computation of a simple $\xi$-dependent tensor, where the latter can  also be neglected in the simplest version of our algorithm. We provide analytical formulas for both expressions. In Section~\ref{qc} we present a quantum circuit that allows one to realize the non-Gaussian state Ansatz on a quantum computer. We summarize the results of our work in Section~\ref{conlclusion}. Detailed calculations, as well as mathematical and numerical methods valuable to the general variational method are included in the appendix.  
%--------------------------------------------------------------------------
%--------------------------------------------------------------------------
%--------------------------------------------------------------------------
%--------------------------------------------------------------------------
\section{Generalized Wick theorem\label{expectation_values}}
This section gives explicit formulas for computing expectation values of products of fermionic operators. More precisely, we will be presenting closed expressions for $\braket{\Psi|\hat c_{j_1}^\dag\cdots \hat c_{j_a}^\dag \hat c_{k_1}\cdots \hat c_{k_b}|\Psi}$, where $j_i,k_i\in \{1,\dots,N_f\}$, $a,b\in\{1,\dots,N_f\}$ and $\ket{\Psi}$ represents a quantum state which is of the form described in Eq.~\eqref{p2}. Due to parity conservation, we require $a+b=2p$, where $p\leq N_f$ is a non-negative integer. We will first discuss the simple case where $\omega=\mathbb 0_{N_f}$ and then move to the general case $\omega\neq\mathbb 0_{N_f}$ in Sections~\ref{part1}-\ref{part2}.

We now consider the case $\omega=\mathbb 0_{N_f}$, where we have $\ket{\Psi}=\ket{\Psi_{\text{GS}}}$ and define  
\begin{align}
    \mathcal A_{j_1^\dag \dots j_a^\dag k_1\dots k_b}^{[0]}=\braket{\Psi_{\text{GS}}|\hat c_{j_1}^\dag\cdots \hat c_{j_a}^\dag \hat c_{k_1}\cdots \hat c_{k_b} |\Psi_{\text{GS}}}.\label{wick2}
\end{align}
Note, that due to normalization, we have $\mathcal A^{[0]}=\braket{\Psi_{\text{GS}}|\Psi_{\text{GS}}}=1$. It is well known that one can compute the above expectation value analytically by means of Wick's theorem \cite{bravyi2005lagrangian},\footnote{Wick's theorem can also be stated for products of Majorana operators as defined in Eqs.~\eqref{ma1}-\eqref{ma2},
$$\braket{\Psi_{\text{GS}}|\hat A_{i_1}\hat A_{i_2}\cdots\hat A_{i_{2p}}|\Psi_{\text{GS}}}=(-i)^p\text{Pf}\left(\left.\Gamma_m\right|_{i_1,i_2,\dots,i_{2p}}\right)$$
where  $\text{Pf}(\left.\Gamma_m\right|_{i_1,i_2,\dots,i_{2p}})$ denotes the Pfaffian of the submatrix $\left.\Gamma_m\right|_{i_1,i_2,\dots,i_{2p}}$ which is obtained by only considering the rows and columns $i_1,...,i_{2p}$ of the skew-symmetric matrix $\Gamma_m$ introduced in Eq.~\eqref{p8} \cite{kraus2009quantum}.}
\begin{align}
\mathcal A_{j_1^\dag \dots j_a^\dag k_1\dots k_b}^{[0]} = \sum_{\pi\in \Pi(j_1^\dag,\dots,j_a^\dag,k_1,\dots,k_b)} \text{sgn}(\pi)\mathcal A_{\pi(1)}^{[0]}\mathcal A_{\pi(2)}^{[0]}\cdots \mathcal A_{\pi(p)}^{[0]}.\label{wick1}
\end{align}
We mapped the original string $(j_1^\dag,\dots,j_a^\dag,k_1,\dots,k_b)\rightarrow(i_1,\dots,i_a,i_{a+1},i_{2p})$ to a string of integers $i_j\in\{1,\dots,2p\}$, where the first $a$ elements $(i_1,\dots,i_a)$ correspond to  $(j_1^\dag,\dots,j_a^\dag)$ and the remaining $b$ elements $(i_{a+1},\dots,i_{2p})$ correspond to $(k_1,\dots,k_b)$, in the same order as they appear in the original string. We then define the set of all partitions of the original string into pairs $x_{i_{2\ell-1}}x_{i_{2\ell}}$ taking into account the order,
\begin{align}
    \Pi(j_1^\dag,\dots,j_a^\dag,k_1,\dots,k_b) =&\left\{\left.  \left(x_{i_1}x_{i_2},x_{i_3}x_{i_4},\dots,x_{i_{2p-1}}x_{i_{2p}}\right)\right|\begin{array}{l}
        i_{2\ell-1}<i_{2\ell}   \\
        i_{2\ell-1}<i_{2\ell+1}
    \end{array}, \ell \in \{1,\dots,p\}, \right.\nonumber\\ &\left.  \ \text{with} \ i_m\in \{1,\dots,N_f\} \ \text{for} \ m \in \{1,\dots,2p\}: x_{i_m} =\left\{\begin{array}{lr}  j_{i_m}^\dag,& \text{if} \ i_m\leq a \\  k_{i_m}, &\text{if} \ i_m> a\end{array}\right\} 
    \right\}.\label{set}
\end{align}
The symbol $\text{sgn}(\pi)$ is the signature of an element $\pi$ of the set defined in Eq.~\eqref{set}, in other words, $\text{sgn}(\pi)$ is identical to $+1$ ($-1$) if the number of nearest neighbour permutations required to recover the original string is even (odd), and $\pi(k)$ with $k\in\{1,\dots,p\}$ outputs the $k$-th pair in $\pi$.\footnote{As a simple example, consider the string $j_1^\dag j_2^\dag j_3 j_4$. Using the definition in Eq.~\eqref{set}, we have the set $\Pi(j_1^\dag j_2^\dag j_3 j_4)=\{\pi_1,\pi_2,\pi_3\}$ with elements $\pi_1=(j_1^\dag j_2^\dag, j_3j_4)$, $\pi_2=(j_1^\dag j_3, j_2^\dag j_4)$ and $\pi_3=(j_1^\dag j_4 ,j_2^\dag j_3)$, with respective signatures $\text{sgn}(\pi_1)=1$, $\text{sgn}(\pi_2)=-1$, and $\text{sgn}(\pi_3)=1$. In particular, one can single-out an individual pair through e.g. $\pi_1(1)=j_1^\dag j_2^\dag$, or $\pi_3(2)=j_2^\dag j_4$. Using Eq.~\eqref{wick1}, the corresponding expectation value is then given by 
$$\mathcal A^{[0]}_{j_1^\dag j_2^\dag j_3 j_4}=\mathcal A^{[0]}_{j_1^\dag j_2^\dag}\mathcal A^{[0]}_{j_3j_4}-\mathcal A^{[0]}_{j_1^\dag j_3}\mathcal A^{[0]}_{j_2^\dag j_4}+\mathcal A^{[0]}_{j_1^\dag j_4}\mathcal A^{[0]}_{j_2^\dag j_3}.$$} The terms on the right-hand side of Eq.~\eqref{wick1} depend on the values of the pairs $x_{i_{2\ell-1}}x_{i_{2\ell}}$ and the fermionic Gaussian state, and are of the following form
\begin{align}
\mathcal A_{p^\dag q}^{[0]}=&\frac{i}{4}\left(\mathcal G^{[0]}\right)_{p,q}^{+-},\label{wick3}\\
\mathcal A_{p^\dag q^\dag}^{[0]}=&\frac{i}{4}\left(\mathcal G^{[0]}\right)_{p,q}^{++},\label{wick4}\\
\mathcal A_{p q}^{[0]}=&\frac{i}{4}\left(\mathcal G^{[0]}\right)_{p,q}^{--},  \label{wick5}
\end{align}
where 
\begin{align}
\mathcal G^{[0]}=\Gamma_m+\Upsilon,\label{wick8}    
\end{align}
and $\Upsilon = \sigma\otimes \mathds 1_{N_f}$ with 
\begin{align}
\sigma =     \begin{pmatrix}
0&1\\-1&0
\end{pmatrix}.\label{p510}
\end{align}
Here, $\Gamma_m$ denotes the real and anti-symmetric \textit{covariance matrix} whose elements are given by
\begin{align}
\left(\Gamma_m\right)_{kl} = \frac{i}{2}\braket{\Psi_{\text{GS}}|[\hat A_k,\hat A_l]|\Psi_{\text{GS}}}\label{p8}.
\end{align}
 The expressions on the right-hand side of Eqs.~\eqref{wick3}-\eqref{wick5} can be computed as follows. For $p,q\in \{1,\dots,N_f\}$ we write
\begin{align}
\begin{pmatrix}
\mathbf 1_p\\\mathbf i_q
\end{pmatrix}=\left(\dots,0_{p-1},1_p,0_{p+1},\dots,0_{N_f+q-1},i_{N_f+q},0_{N_f+q+1},\dots\right)^T.
\end{align}
This allows us to abbreviate particular sums of matrix elements with respect to a $(2N_f\times 2N_f)$-matrix $M$ through
\begin{align}
M^{+-}_{p,q} =&\begin{pmatrix}
\mathbf 1_q^T,\mathbf i_q^T
\end{pmatrix}M\begin{pmatrix}
\mathbf 1_p\\
-\mathbf i_p
\end{pmatrix},\label{p490}\\
M^{-+}_{p,q} =&\begin{pmatrix}
\mathbf 1_q^T,-\mathbf i_q^T
\end{pmatrix}M\begin{pmatrix}
\mathbf 1_p\\
\mathbf i_p
\end{pmatrix},\label{p596}\\
M^{++}_{p,q} =&\begin{pmatrix}
\mathbf 1_q^T,-\mathbf i_q^T
\end{pmatrix}M\begin{pmatrix}
\mathbf 1_p\\
-\mathbf i_p
\end{pmatrix},\label{p491}\\
M^{--}_{p,q} =&\begin{pmatrix}
\mathbf 1_q^T,\mathbf i_q^T
\end{pmatrix}M\begin{pmatrix}
\mathbf 1_p\\
\mathbf i_p
\end{pmatrix}.\label{p492}
\end{align}
Note the unusual index ordering in the definitions above, which is used exclusively for expressions containing the superscripts $++,+-,-+,--$. 

The efficiency of the computation of Eq.~\eqref{wick1} is owed to the fact, that the evaluation of the expectation value in the $(2^{N_f}\times 2^{N_f})$-dimensional Hilbert space can be broken down to the evaluation of matrix elements of the $(2N_f\times 2N_f)$-covariance matrix $\Gamma_m$. Here, $\Gamma_m$ in Eq.~\eqref{p8} can be directly calculated from $\xi$ given in Eq.~\eqref{p6} through  $\Gamma_m=-U_m\Upsilon U_m^T$, where $U_m=e^{i\xi}$. This allows us to express the set of variational parameters describing the non-Gaussian state Ansatz of Eq.~\eqref{p2} either in terms of $\{\xi,\omega\}$ or $\{\Gamma_m,\omega\}$.

In this work, we are interested in the general case $\omega\neq \mathbb 0_{N_f}$, where $\ket{\Psi}=\ket{\Psi_{\text{NGS}}}$. Since $\hat U_{\text{FA}}$ is an exponential of a \textit{quartic} polynomial of fermionic operators, $\ket{\Psi_{\text{NGS}}}$ is no longer a fermionic Gaussian state, since the ladder are generated by exponentials of a \textit{quadratic} polynomial of fermionic operators. As a consequence, we require a method that allows us to compute $\braket{\Psi_{\text{NGS}}|\hat c_{j_1}^\dag\cdots \hat c_{j_a}^\dag \hat c_{k_1}\cdots \hat c_{k_b}|\Psi_{\text{NGS}}}$ efficiently. This was accomplished by Shi \textit{et al.} in \cite{SHI2018245}, who provided a generating function of Grassmann variables that allows one to compute said expectation values efficiently by taking Grassmann derivatives on the generating function. We use this representation to derive a generalized Wick theorem, that agrees in the special case $\omega=\mathbb 0_{N_f}$ with the usual Wick theorem. 
%--------------------------------------------------------------------------
\subsection{Implicit formula\label{part1}}
This subsection states the main results of  \cite{SHI2018245} regarding the computation of expectation values with respect to fermionic non-Gaussian states. The unitary operator $\hat U_{\text{FA}}$ transforms a fermionic annihilation operator as follows,
\begin{align}
\hat U_{\text{FA}}^\dag \hat c_k \hat U_{\text{FA}} = e^{i\sum_{j}\omega_{jk}\hat n_j}\hat c_k, \label{p352}
\end{align}
attaching a phase depending on particles in modes $j$ to the particle in mode $k$, reminiscent of the Flux Attachment (FA) procedure known from composite fermion theory \cite{jain2007composite}. One can commute the single fermionic operator on the right-hand side of Eq.~\eqref{p352} past the exponential term through $e^{i\sum_j \omega_{jq}\hat n_j}\hat c_p=e^{-i\omega_{pq}}\hat c_pe^{i\sum_{j}\omega_{jq}\hat n_j}\label{p46}$, which allows us to rewrite $\braket{\Psi_{\text{NGS}}|\hat c_{j_1}^\dag\cdots \hat c_{j_a}^\dag \hat c_{k_1}\cdots \hat c_{k_b}|\Psi_{\text{NGS}}}$ as expectation values of a particular type of fermionic operator with respect to a fermionic Gaussian state,
\begin{align}
\mathcal A_{j_1^\dag\dots j_a^\dag k_1\dots k_b}^{[\tilde \alpha]}=& \left\langle \Psi_{\text{GS}}\right| e^{i\sum_j\tilde \alpha(j)\hat n_j} \hat c_{j_1}^\dag\cdots \hat c_{j_a}^\dag \hat c_{k_1}\cdots \hat c_{k_b}\left|\Psi_{\text{GS}} \right\rangle.\label{p228}
\end{align}
Here, $\tilde\alpha=(\tilde\alpha(1),\dots,\tilde\alpha(N_f))$ is a vector of length $N_f$, with $\tilde\alpha(j)\in\mathds R$, that implicitly depend on the non-Gaussian variational parameters $\tilde\alpha=\tilde\alpha(\omega)$. In particular, for $\omega=\mathbb 0_{N_f}$, we have $\tilde\alpha = (0,\dots,0)$ and Eq.~\eqref{p228} reduces to Eq.~\eqref{wick2}. Shi \textit{et al.} showed that Eq.~\eqref{p228} can be computed through
\begin{align}
\mathcal A_{j_1^\dag\dots j_a^\dag k_1\dots k_b}^{[\tilde\alpha]} = \mathcal A^{[\tilde\alpha]}\mathcal F_{J_{j_1}\dots J_{j_a} J_{k_1}^*\dots J_{k_b}^*} e^{X^{[\tilde\alpha]}},\label{p818}
\end{align}
where $\mathcal A^{[\tilde\alpha]}$ is a complex coefficient defined in Eq.~\eqref{p819}, $X^{[\tilde\alpha]}$ is a quadratic polynomial of Grassmann variables that depends on $\tilde\alpha$ (indicated by the superscript $[\tilde\alpha]$) defined in Eq.~\eqref{p817}, and 
\begin{align}
&\mathcal F_{J_{j_1},\dots,J_{k_a},J_{k_1}^*,\dots J_{k_b}^*}f(J_{1},\dots,J_{N_f},J_{1}^*,\dots,J_{N_f}^*)\nonumber\\=&\lim\limits_{J_{j_1}\rightarrow 0}\cdots \lim\limits_{J_{j_a}\rightarrow 0}\lim\limits_{J_{k_1}^*\rightarrow 0}\cdots \lim\limits_{J_{k_b}^*\rightarrow 0}\frac{d}{d J_{j_1}}\cdots \frac{d}{d J_{j_a}}\frac{d}{d J_{k_1}^*}\cdots \frac{d}{d J_{k_b}^*}f(J_{1},\dots,J_{N_f},J_{1}^*,\dots,J_{N_f}^*)\label{p810}
\end{align} 
defines a left-derivative operator of Grassmann variables $J_i$ and the conjugated Grassmann variables $J_i^*$ acting on some differentiable function of Grassmann variables $f(J_{1},\dots,J_{N_f},J_{1}^*,\dots,J_{N_f}^*)$. Grassmann variables are anticommuting, in that they satisfy $\{J_i,J_j\}=\{J_i^*,J_j^*\}=\{J_i,J_j^*\}=0$ \cite{cahill1999density}. On the right-hand side of Eq.~\eqref{p818}, we define the quadratic polynomial of Grassmann variables 
\begin{align}
X^{[\tilde\alpha]}=&\frac{1}{2i}\boldsymbol{\vec{J}}^{\ \dag}{\Sigma_F^{[\tilde\alpha]}}^\dag\sqrt{1-e^{i\tilde\alpha}}\left({\Gamma_F^{[\tilde\alpha]}}\right)^{-1}\sqrt{1-e^{i\tilde\alpha}} \Sigma_F^{[\tilde\alpha]}\boldsymbol{\vec{J}}+\sum_{k=1}^{N_f} \frac{J_k^*J_k}{1-e^{-i\tilde\alpha(k)}},\label{p817}
\end{align}
whose matrix exponential $e^{X^{[\tilde\alpha]}}$ is the generating function  and $\boldsymbol{\vec{J}}=(J_1,\dots,J_{N_f},J_1^*,\dots,J_{N_f}^*)^T$. We define the complex-valued expression
\begin{align}
\mathcal A^{[\tilde\alpha]}=\left\langle\Psi_{\text{GS}}\right| e^{i\sum_j\tilde \alpha(j)\hat n_j}\left|\Psi_{\text{GS}} \right\rangle=s_{N_f}\left(\frac{1}{2}\right)^{N_f} \text{Pf}\left(\Gamma_F^{[\tilde\alpha]}\right)\label{p819}.  
\end{align}
Here, $s_{N_f}$ is a sign factor that depends on the parity of the number of spin-orbitals $N_f$,
\begin{align}
s_{N_f}=\begin{cases}
(-1)^{\tfrac{{N_f}}{2}}, &\ \text{if ${N_f}$ is even,}\\
(-1)^{\tfrac{{N_f}+1}{2}}, &\ \text{if ${N_f}$ is odd,}
\end{cases}\label{p816}
\end{align}
while 
\begin{align}
\Gamma_F^{[\tilde\alpha]}=&\sqrt{1-e^{i\tilde\alpha}}\Gamma_m \sqrt{1-e^{i\tilde\alpha}}-\sigma\otimes \text{diag}\left(1+e^{i\tilde\alpha}\right).\label{p814}
\end{align}
We define
\begin{align}
\sqrt{1\mp e^{i\tilde\alpha}} =&
\mathds 1_2\otimes \begin{pmatrix}
\sqrt{1\mp e^{i\tilde \alpha(1)}}&&\\
&\ddots&\\
&&\sqrt{1\mp e^{i\tilde \alpha{(N_f)}}}\\
\end{pmatrix},\label{p813}
\end{align}
and 
\begin{align}
\text{diag}\left(1\pm e^{i\tilde \alpha}\right)=&\begin{pmatrix}
1 \pm e^{i\tilde \alpha(1)}&&\\
&\ddots&\\
&&1\pm e^{i\tilde \alpha{(N_f)}}
\end{pmatrix}.\label{p815}
\end{align}
For readability, zeros are not displayed and we use $\mathds 1_k$ to denote the $(k\times k)$-identity matrix. In Eq.~\eqref{p817}, we further define the column vector of Grassmann variables of length $2N_f$, \redsout{$\boldsymbol{\vec{J}}=(J_1,\dots,J_{N_f},J_1^*,\dots,J_{N_f}^*)^T$,} as well as the $(2N_f\times 2N_f)$-matrix
\begin{align}
&\Sigma_F^{[\tilde\alpha]}=\begin{pmatrix}
\tfrac{e^{i\tilde\alpha(1)}}{1-e^{i\tilde\alpha(1)}}  &  &  & \tfrac{-1}{1-e^{i\tilde\alpha(1)}} &  & \\
&\ddots&&&\ddots&\\
& & \tfrac{e^{i\tilde\alpha{(N_f)}}}{1-e^{i\tilde\alpha{(N_f)}}} & && \tfrac{-1}{1-e^{i\tilde\alpha{(N_f)}}}\\
\tfrac{-ie^{i\tilde\alpha(1)}}{1-e^{i\tilde\alpha(1)}} &  & & \tfrac{-i}{1-e^{i\tilde\alpha(1)}} & &\\
&\ddots&&&\ddots&\\
&& \tfrac{-ie^{i\tilde\alpha{(N_f)}}}{1-e^{i\tilde\alpha{(N_f)}}}&&& \tfrac{-i}{1-e^{i\tilde\alpha{(N_f)}}}
\end{pmatrix}.\label{p812}
\end{align} 
%--------------------------------------------------------------------------
\subsection{Explicit formula\label{part2}}
Since for Grassmann variables, left differentiation is identical to integration and $X^{[\tilde\alpha]}$ in Eq.~\eqref{p817} is a quadratic polynomial in the Grassmann variables, the expression $\mathcal F_{J_{j_1}\dots J_{j_a}J_{k_1}^*\dots J_{k_b}^*} e^{X^{[\tilde\alpha]}}$ on the right-hand side of Eq.~\eqref{p818} can be written as a Grassmann integral over the exponential of the quadratic polynomial of Grassmann variables $X^{[\tilde\alpha]}$ \cite{cahill1999density,berezin1996method}. This can be evaluated to give
\begin{align}
\frac{\mathcal A_{j_1^\dag \dots j_a^\dag k_1\dots k_b}^{[\tilde \alpha]}}{\mathcal A^{[\tilde\alpha]}} =  \sum_{\pi\in \Pi(j_1^\dag,\dots,j_a^\dag,k_1,\dots,k_b)} \text{sgn}(\pi)\frac{\mathcal A_{\pi(1)}^{[\tilde\alpha]}}{\mathcal A^{[\tilde\alpha]}}\frac{\mathcal A_{\pi(2)}^{[\tilde\alpha]}}{\mathcal A^{[\tilde\alpha]}}\cdots \frac{\mathcal A_{\pi(p)}^{[\tilde\alpha]}}{\mathcal A^{[\tilde\alpha]}},\label{p360}
\end{align}
which is a direct consequence of Wick's theorem, see e.g. chapter 2.3 in \cite{Negele1998Quantum}. In fact, Eq.~\eqref{p360} can be seen to be identical to Eq.~\eqref{p818}, by noting that we can write $\braket{\Psi_{\text{GS}}|e^{i\sum_j\tilde \alpha(j)\hat n_j} \hat c_{j_1}^\dag\cdots \hat c_{j_a}^\dag \hat c_{k_1}\cdots \hat c_{k_b}|\Psi_{\text{GS}}}=\text{tr}\left( \hat O_{\text{GS}}^{[\tilde\alpha]} \hat c_{j_1}^\dag\cdots \hat c_{j_a}^\dag \hat c_{k_1}\cdots \hat c_{k_b} \right)$, where $\hat O_{\text{GS}}^{[\tilde\alpha]}=\ket{\Psi_{\text{GS}}}\bra{\Psi_{\text{GS}}} e^{i\sum_j\tilde\alpha(j)\hat n_j}$ is a Gaussian operator with trace $\text{tr}\left(\hat O_{\text{GS}}^{[\tilde\alpha]}\right)=\mathcal A^{[\tilde\alpha]}$. It is known \cite{Corney2006GaussianPR}, that   $\text{tr}\left( \hat O_{\text{GS}}^{[\tilde\alpha]} \hat c_{j_1}^\dag\cdots \hat c_{j_a}^\dag \hat c_{k_1}\cdots \hat c_{k_b} \right)=\text{tr}\left(\hat O_{\text{GS}}^{[\tilde\alpha]}\right)\times\text{``Wick's theorem''}$, where ``Wick's theorem'' is identical to the right-hand side of Eq.~\eqref{p360}---in other words, it is  constructed from the two-point correlation function $A_{jk}=\text{tr}\left(\hat O_{\text{GS}}^{[\tilde\alpha]}\hat x_j\hat x_k\right)/\mathcal A^{\tilde\alpha}$, where $\hat x_j\neq \hat x_k$ can be any of the fermionic operators $\hat c_{j_1}^\dag,\dots,\hat c_{j_a}^\dag,\hat c_{k_1},\dots,\hat c_{k_b}$. Note, that Eq.~\eqref{wick1} is a special case of \eqref{p360}, since $\mathcal A^{[0]}=1$.

In order to be able to compute Eq.~\eqref{p360}, one requires explicit expressions for $\mathcal A_{p^\dag q}^{[\tilde\alpha]}$, $\mathcal A_{p^\dag q^\dag}^{[\tilde\alpha]}$, and $\mathcal A_{p q}^{[\tilde\alpha]}$, which are generalizations of Eqs.~\eqref{wick3}-\eqref{wick5}. While $\mathcal A^{[\tilde\alpha]}$ and $\mathcal A_{p^\dag q}^{[\tilde\alpha]}$ are given in  \cite{SHI2018245}, we derive explicit formulas for the missing two expressions in Appendix~\ref{expectation}, where we also present an alternative derivation for the exemplary expression $\mathcal A^{[\tilde\alpha]}_{p^\dag q^\dag rs}$ using the implicit formula in Eq.~\eqref{p818}, which is shown to be identical to the expression obtained from Eq.~\eqref{p360}. The findings are summarized below,
\begin{align}
\mathcal A_{p^\dag q}^{[\tilde\alpha]}=&\frac{i}{4}e^{i\tilde\alpha(p)}\mathcal A^{[\tilde\alpha]}\left(\mathcal G^{[\tilde\alpha]}\right)_{p,q}^{+-},\label{p366}\\
\mathcal A_{p^\dag q^\dag}^{[\tilde\alpha]}=&\frac{i}{4}e^{i(\tilde\alpha(p)+\tilde\alpha(q))}\mathcal A^{[\tilde\alpha]}\left(\mathcal G^{[\tilde\alpha]}\right)_{p,q}^{++},\label{p367}\\
\mathcal A_{p q}^{[\tilde\alpha]}=&\frac{i}{4}\mathcal A^{[\tilde\alpha]}\left(\mathcal G^{[\tilde\alpha]}\right)_{p,q}^{--}.  \label{p368}
\end{align}
Here, $\tilde\alpha(p)$ is the $p$-th entry of the vector $\tilde\alpha$. Following \cite{SHI2018245}, we define the skew-symmetric matrix
\begin{align}
\mathcal G^{[\tilde\alpha]} =\frac{\Gamma_m+\Upsilon}{\mathds 1_{2N_f}+\tfrac{1}{2}\left(1-e^{i\tilde\alpha}\right)\left(\Upsilon\Gamma_m-\mathds 1_{2N_f}\right)}\label{p365},
\end{align}
which reduces to Eq.~\eqref{wick8} for $\tilde\alpha=(0,\dots,0)$. Here,  $(1-e^{i\tilde\alpha})=(\sqrt{1-e^{i\tilde\alpha}})^2$ and we use the short-hand notation $\frac{A}{B}=AB^{-1}$ for two equal-sized matrices $A$ and $B$, provided that the latter is invertible. 

In addition to the explicit formulas in Eqs.~\eqref{p360} and \eqref{p367}-\eqref{p368}, we found that the inverse appearing in Eq.~\eqref{p365} can be evaluated without having to perform the actual, numerically costly matrix inversion using the method presented in \cite{miller1981inverse}.\footnote{In order to circumvent having to perform the matrix inversion in Eq.~\eqref{p365}, we rewrite $\mathcal G^{[\tilde\alpha]}=-\Upsilon[A+B]^{-1}$, with $A= -\left(\Gamma_m+\Upsilon\right)^{-1}\Upsilon$ and  $B=\frac{1}{2}\left(1-e^{i\tilde\alpha}\right)$. We find $A^{-1}=\Upsilon(\Gamma_m+\Upsilon)$. We give a decomposition $B=\sum_{k=1}^{r_B}B_{i_k}=\sum_{k=1}^{r_B}\tilde\beta_k\ket{i_k}\bra{i_k}$ into rank 1 matrices $B_k$, where we defined the element $i_k$ of a list $[i_1,....,i_{r_B}]$ containing all non-zero index entries of the diagonal terms of $B$,  $\tilde\beta_k=\frac{1}{2}(1-e^{i\tilde\alpha(i_k)})$, and $r_B\leq 2N_{f}$ is identical to the rank of $B$. We define $C_{\ell+1}=A+\sum_{k=1}^\ell B_{i_k}$ with $\ell\leq r_B$, where $C_{r_B+1}=A+B$ and $C_1=A$. The iterative procedure of  \cite{miller1981inverse} is given by $C_{\ell+1}^{-1}=C_\ell^{-1}-g_\ell C_\ell^{-1}B_{i_\ell}C_\ell^{-1}$, with $g_{\ell}=1/(1+\text{tr}(C_\ell^{-1}B_{i_\ell}))$. This strategy can also be used for other instances within the variational method, see Appendix~\ref{H_m}.} This has the potential to significantly speedup numerical computations, as in each time-step one has to compute $\mathcal G^{[\tilde\alpha]}$ for each distinct $\tilde\alpha$ and this method allows one to separate the effect of the non-Gaussian parameters $\omega$ from the Gaussian parameters in $\Gamma_m$.  The following section will show how the results derived here can be used to compute the energy expectation value of a general quantum many-body Hamiltonian. 
%--------------------------------------------------------------------------
%--------------------------------------------------------------------------
%--------------------------------------------------------------------------
%--------------------------------------------------------------------------
\section{Energy expectation value\label{qmb}}
In this section, we derive an analytical formula for the energy expectation value in Eq.~\eqref{p350} in terms of the expression of Eq.~\eqref{p360}. We consider interacting fermionic systems whose non-relativistic dynamics are governed by the Hamiltonian\footnote{We want to stress that the variational method can also be applied to more general parity-conserving Hamiltonians, which could lead to larger (even) products of fermionic operators appearing in Eq.~\eqref{p58}.}
\begin{align}
\hat H=\sum_{p,q=1}^{N_f} f_{pq}\hat c_p^\dag \hat c_q + \frac{1}{2}\sum_{p,q,r,s=1}^{N_f}h_{pqrs}\hat c_p^\dag \hat c_q^\dag \hat c_r \hat c_s,\label{p58}
\end{align}
where $f_{pq}$ and $h_{pqrs}$ are the one- and two-body matrix elements as defined in  \cite{helgaker2014molecular}, where the latter possess the symmetries $h_{pqrs}=-h_{qprs}=-h_{pqsr}=h_{qpsr}$ and $h_{pqrs}={h_{srqp}}$. For the last condition, we assume that values of the resulting two-body integrals are real-valued. We rewrite the energy expectation value in Eq.~\eqref{p350} for the Hamiltonian defined in Eq.~\eqref{p58} in terms of the expectation value of the rotated Hamiltonian operator
\begin{align}
\hat H_{\text{FA}} = \hat U_{\text{FA}}^\dag \hat H \hat U_{\text{FA}} \label{p452}
\end{align}
with respect to a fermionic Gaussian state, so that $E(\xi,\omega)=\braket{\Psi_{\text{GS}}(\xi)|\hat H_{\text{FA}}(\omega)|\Psi_{\text{GS}}(\xi)}$. The rotated Hamiltonian in Eq.~\eqref{p452} can be written as $\hat H_{\text{FA}}=\hat H^{(1)}_{\text{FA}}+\hat H^{(2)}_{\text{FA}}$, where
\begin{align}
\hat H^{(1)}_{\text{FA}}=&\sum_{p,q=1}^{N_f}f^{\text{FA}}_{pq} e^{i\sum_k\alpha_{pq}^{\text{FA}}(k) \hat n_{k}}  \hat c_p^\dag \hat c_q,\label{p581}\\
\hat H^{(2)}_{\text{FA}}=&\frac{1}{2}\sum_{p,q,r,s=1}^{N_f}h_{pqrs}^{\text{FA}}e^{i\sum_k \beta_{pqrs}^{\text{FA}}(k) \hat n_{k}}  \hat c_p^\dag \hat c_q^\dag \hat c_r \hat c_s,\label{p582}
\end{align}
are the rotated one- and two-body Hamiltonian terms, and $k=1,\dots,N_f$. We defined  $f^{\text{FA}}_{pq}= f_{pq}e^{-i\alpha_{pq}^{\text{FA}}(p)}$, $h_{pqrs}^{\text{FA}}=h_{pqrs}e^{-i\left(\beta_{pqrs}^{\text{FA}}(p)+\beta_{pqrs}^{\text{FA}}(q)\right)}$, $\alpha_{pq}^{\text{FA}}(k)=\omega_{kq}-\omega_{kp}$, and $\beta_{pqrs}^{\text{FA}}(k)=\alpha_{ps}^{\text{FA}}(k)+\alpha_{qr}^{\text{FA}}(k)$. It is important to note, that the operators of the rotated Hamiltonian  in Eqs.~\eqref{p581}-\eqref{p582} differ considerably from the operators in Eq.~\eqref{p58}, as they are no longer just quadratic and quartic polynomials of fermionic operators, but polynomials of up to order $N_f$. However,  the total number of terms appearing in $\hat H_{\text{FA}}$ is identical to the number of terms in the system Hamiltonian $\hat H$ in Eq.~\eqref{p58}. Using the above results, Eq.~\eqref{p350} can be written as $E\left(\xi,\omega\right) = E_1(\xi,\omega)+E_2(\xi,\omega)$,
where the one- and two-body energy expectation values are given by
\begin{align}
E_1(\xi,\omega) =& \sum_{p,q}f^{\text{FA}}_{pq}\mathcal A_{p^\dag q}^{[\alpha_{pq}^{\text{FA}}]}\label{p832}\\
E_2(\xi,\omega) =&\frac{1}{2}\sum_{p,q,r,s}h_{pqrs}^{\text{FA}}\mathcal A_{p^\dag q^\dag rs}^{[\beta_{pqrs}^{\text{FA}}]},\label{p833}
\end{align}
which depend implicitly on the variational parameters $\{\xi,\omega\}$ (or $\{\Gamma_m,\omega\}$) in both, the Hamiltonian coefficients $f^{\text{FA}}_{pq}$ and $h_{pqrs}^{\text{FA}}$, as well as in the expectation values $\mathcal A_{p^\dag q}^{[\alpha_{pq}^{\text{FA}}]}$ and $\mathcal A_{p^\dag q^\dag rs}^{[\beta_{pqrs}^{\text{FA}}]}$. This motivates how expectation values of Eq.~\eqref{p228} appear as a consequence of the unitary transformation of Eq.~\eqref{p352} due to the presence of the flux attachment operator $\hat U_{\text{FA}}$ in the variational state Ansatz.\footnote{In  \cite{SHI2018245}, it is suggested to use a variational state Ansatz with more degrees of freedom when studying e.g. fractional quantum Hall systems. The Ansatz proposed by Shi \textit{et al.} adds an additional fermionic Gaussian operator to the left of Eq.~\eqref{p2}, such that $\ket{\Psi_{\text{NGS}}(\xi,\omega,\xi')}=\hat U_{\text{GS}}(\xi')\hat U_{\text{FA}}(\omega)\hat U_{\text{GS}}(\xi)\ket{0}$. Since the transformation defined by the action of $\hat U_{\text{GS}}$ on a Majorana operator is described by the following linear operation \cite{bravyi2005lagrangian}
\begin{align*}
\hat U_{\text{GS}}^\dag \hat A_j \hat U_{\text{GS}} = \sum_{k=1}^{2N_f}\left(U_m\right)_{jk}\hat A_k,
\end{align*}
this extended variational Ansatz $\ket{\Psi_{\text{NGS}}(\xi,\omega,\xi')}$ essentially rotates the Hamiltonian in Eq.~\eqref{p58} into a new basis which depends on $\xi'$, and its only added difficulty is the requirement of an additional updating rule for the time evolution of the second set of Gaussian variational parameters $\xi'$, e.g. through Eq.~\eqref{p410}. We will leave a treatment of this more complex Ansatz to future works.}

The closed analytical formulas derived in Eqs.~\eqref{p832}-\eqref{p833} allow one to compute the energy expectation value of Eq.~\eqref{p350} for any parity conserving physical Hamiltonian through Eq.~\eqref{p360}. By employing a constrained optimization algorithm, one can find a local minimum of the rotated many-body Hamiltonian $\hat H_{\text{FA}}$ as defined in Eq.~\eqref{p452} within the family of fermionic Gaussian states. The role of the optimization algorithm is to provide an updating rule for the variational parameters (here, $\Gamma_m$ and $\omega$), while still outputting physical states which do not violate certain symmetry properties, such as the skew-symmetry (symmetry) of $\Gamma_m$ ($\omega$), or the purity of the Gaussian state $\Gamma_m^2=-\mathds 1_{2N_f}$. In the following section, we will derive an  optimization strategy for  the non-Gaussian variational parameters $\omega$ that offers a simple alternative to the imaginary-time based approach presented in  \cite{SHI2018245} and does not violate any symmetry constraints. The algorithm is based on the principle of ensuring a monotonic decrease of the energy in each optimization step. 
%--------------------------------------------------------------------
%--------------------------------------------------------------------
%--------------------------------------------------------------------
%--------------------------------------------------------------------
\section{A hybrid minimization algorithm\label{hybrid}}
In this section, we will derive an alternative optimization strategy for the non-Gaussian variational parameters, that, in its simplest form, only requires the evaluation of the energy gradient with respect to $\omega$. This section is structured as follows. In Section~\ref{ite_shi} we present the imaginary time evolution-based optimization strategy derived by Shi \textit{et al.} \cite{SHI2018245}. We compute how the energy expectation value of Eq.~\eqref{p350} evolves in time for the non-Gaussian state Ansatz of Eq.~\eqref{p2} in Section~\ref{te_en_exp} and show how this can be used as the basis for a simplified hybrid imaginary-time-gradient-descent optimization algorithm in Section~\ref{hitgd_ansatz}.
%--------------------------------------------------------------------
\subsection{Optimization algorithm based on imaginary time evolution\label{ite_shi}}
Let us first review the optimization strategy employed by Shi \textit{et al.} \cite{SHI2018245}. It is  based on imaginary time evolution of the variational state, 
\begin{align}
d_\tau \ket{\Psi_{\text{NGS}}} =-(\hat H-E)\ket{\Psi_{\text{NGS}}}\label{p434}
\end{align}
where $E$ is the energy expectation value of Eq.~\eqref{p350} and $d_\tau = \frac{d}{d\tau}$ denotes the partial derivative with respect to the imaginary time $\tau$. We define the non-Gaussian Ansatz-dependent operator $\hat O=\tfrac{i}{2}\sum_{i\neq j}\left(d_\tau\omega\right)_{ij}:\hat n_i\hat n_j:$, the mean-field matrix of the rotated Hamiltonian $H_{\text{FA}_m}=4\frac{d}{d\Gamma_m}\braket{\Psi_{\text{GS}}|\hat H_{\text{FA}}|\Psi_{\text{GS}}}$, and similarly the mean-field matrix of the non-Gaussian state Ansatz $O_m=4\frac{d}{d\Gamma_m}\braket{\Psi_{\text{GS}}|\hat O|\Psi_{\text{GS}}}$. Here, $\frac{d}{d\Gamma_m}$ denotes the derivative with respect to a structured matrix (in this case, the skew-symmetric matrix $\Gamma_m$). We present explicit expressions for $O_m$ and ${H_\text{FA}}_m$ in Appendix~\ref{mf_var_method}. By performing a normal-order expansion of the resulting expressions on the left- and right-hand side of Eq.~\eqref{p434}, one finds the imaginary time equations of motion for the Gaussian variational parameters to be \cite{SHI2018245}
\begin{align}
    d_\tau \Gamma_m = -{H_{\text{FA}}}_m-\Gamma_m {H_{\text{FA}}}_m\Gamma_m+i\left[\Gamma_m,O_m\right].\label{p105}
\end{align}
The time evolution of the non-Gaussian parameters $d_\tau\omega$ on the other hand \redsout{can} can be computed through the following general principle, which follows from Eq.~\eqref{p434}. For any variational parameter $\theta_P$ which can either represent a non-Gaussian parameter or a Gaussian parameter, the imaginary time equations of motion lead to
\begin{align}
    d_\tau \theta_P = \sum_{Q=1}^{N_{\text{var}}}\left(G^{-1}\right)_{PQ}\left\langle \Psi_{Q}|\mathbf R_{\Psi} \right\rangle, \label{p410}
\end{align}
where $N_{\text{var}}$ covers all independent variational parameters, and we defined  $\ket{\mathbf R_{\Psi}}=-(\hat H-E)\ket{\Psi_{\text{NGS}}}$, the tangent vector $\ket{\Psi_P}=d_{\theta_P}\ket{\Psi_{\text{NGS}}}$, as well as the Gram matrix $G_{PQ} = \left\langle\Psi_P|\Psi_Q\right\rangle$. As shown in  \cite{SHI2018245}, the energy decreases monotonically in each time step $\Delta\tau$ if the variational parameters are updated through Eqs.~\eqref{p105} and ~\eqref{p410}. Computing Eq.~\eqref{p410} is efficient, but nevertheless challenging, as it involves computing $\ket{\Psi_P}$, $\ket{\mathbf{R}_{\Psi}}$, as well as generalized Wick expressions for up to eighth order expressions\footnote{Provided the Hamiltonian considered is of the form of Eq.~\eqref{p58}. For Hamiltonians containing larger products of fermionic operators, the expressions will be of even higher order.} resulting from the respective state overlaps, and the (pseudo-)inverse of the Gram matrix. We present explicit formulas for the respective tangent vectors which are required for evaluating Eq.~\eqref{p410} in Appendix~\ref{tangent}. In the following section, we study how the energy changes in time for the non-Gaussian state Ansatz. This will eventually allow us to derive a simple condition that avoids the cumbersome evaluation of Eq.~\eqref{p410}. 
%--------------------------------------------------------------------
\subsection{Time evolution of the energy functional\label{te_en_exp}}
As already mentioned at the end of Section~\ref{qmb}, possessing an analytical expression of the energy functional allows any constrained optimization method to be used to find a local minimum in the high-dimensional energy landscape and thus the set of variational parameters which give an approximate ground state to the system Hamiltonian of Eq.~\eqref{p58}. In the following, we will compute the evolution of the energy as a function of the variational parameters and their respective time-derivatives for the variational states of the form of Eq.~\eqref{p2}. Assuming that the energy depends only on time through $\xi$ (or alternatively $\Gamma_m$) and $\omega$, the time-evolution of the energy expectation value is given by
\begin{align}
d_\tau E(\xi,\omega) =& \frac{1}{4}\sum_{i,j=1}^{2N_f}\left( {H_{\text{FA}}}_m\right)_{ij}d_\tau \left(\Gamma_m\right)_{ij}+\sum_{k,l=1}^{N_f}\frac{dE}{d\omega_{kl}}d_\tau \omega_{kl}.\label{p557}
\end{align}
Using Eq.~\eqref{p105}, Eq.~\eqref{p557} can be written as 
\begin{align}
d_\tau E(\xi,\omega) =&\frac{1}{8}\text{tr}\left(\left(\left[{H_{\text{FA}}}_m,\Gamma_m\right]-iO_m\right)^2\right)+\frac{1}{8}\text{tr}\left(O_m^2\right)+\sum_{k,l=1}^{N_f}\frac{dE}{d\omega_{kl}}d_\tau \omega_{kl}.\label{p556} 
\end{align}
Since $iO_m, \Gamma_m$, and  ${H_{\text{FA}}}_m$ are real-valued skew-symmetric (and therefore anti-symmetric) matrices, the first term on the right-hand side of Eq.~\eqref{p556} contains the trace of the square of a skew-symmetric matrix, which is a symmetric negative semi-definite matrix and thus possesses only non-positive eigenvalues. We therefore have 
\begin{align}
\text{tr}\left(\left(\left[{H_{\text{FA}}}_m,\Gamma_m\right]-iO_m\right)^2\right)\leq 0.\label{p562}
\end{align}
On the other hand, a similar argument cannot be made for the last two terms in Eq.~\eqref{p556}, which means that they may become positive and  the energy will in general not be monotonically decreasing in time. However,  Eq.~\eqref{p556} can serve as a starting point for the construction of alternative optimization approaches, which for instance replace the updating rule for the  non-Gaussian parameter in Eq.~\eqref{p410} with a computationally less demanding Ansatz, while letting the Gaussian parameters continue to evolve under Eq.~\eqref{p105}. We provide an example how the expression in Eq.~\eqref{p556} can inspire simple, possibly computational less demanding optimization algorithms while  guaranteeing a monotonically decreasing energy in each time step in the following subsection. 
%--------------------------------------------------------------------
\subsection{A gradient-descent-based optimization algorithm for the non-Gaussian variational parameters\label{hitgd_ansatz}}
In what follows, we will present our algorithm and defer detailed mathematical derivations to Appendices~\ref{math_deriv}-\ref{en_grad}. A simple choice for evolving the non-Gaussian variational parameters is through a basic gradient-descent Ansatz,
\begin{align}
\left(d_\tau\omega\right)_{ij}=-\frac{1}{c}\frac{dE}{d\omega_{ij}},\label{p724}
\end{align}
where $c\in\mathds R$. As stated in the previous subsection, this choice can in principal cause the energy to increase during a time step. Such a non-monotonic behavior can however be averted by a more carefully designed gradient-based Ansatz, as we show in the following. We choose $d_\tau\omega$ so that the following identity holds,
\begin{align}
    \frac{1}{8}\text{tr}\left(O_m^2\right)+\sum_{k,l=1}^{N_f}\frac{dE}{d\omega_{kl}}\left(d_\tau \omega\right)_{kl}=0.\label{om1}
\end{align}
From Eq.~\eqref{p562} it then automatically follows that the energy monotonically decreases in each time step, see Eq.~\eqref{p556}. We show in Appendix~\ref{math_deriv}, that the first term on the left-hand side in the above equation can be written as a quadratic expression,
\begin{align}
    \text{tr}\left(O_m^2\right) = \sum_{k,l,m,n=1}^{N_f}(d_\tau\omega)_{kl}\mathcal B_{klmn}(d_\tau\omega)_{mn},\label{om2}
\end{align}
where $\mathcal B$ is a $(N_f\times N_f\times N_f\times N_f)$-tensor, which only depends on the covariance matrix $\Gamma_m$. Its entries are given by
\begin{align}
    \mathcal B_{klmn} =&\frac{1}{8}\left[\left(gg^T\right)_{lm}\delta_{nk}+\left(gg^T\right)_{ln}\delta_{mk}+\left(gg^T\right)_{km}\delta_{nl}+\left(gg^T\right)_{kn}\delta_{ml}\right]\nonumber\\&+\frac{1}{8}\left[\left(\mathcal G^{[0]}_{k,l}\right)^2+\left(\mathcal G^{[0]}_{k,N_f+l}\right)^2+\left(\mathcal G^{[0]}_{N_f+k,l}\right)^2+\left(\mathcal G^{[0]}_{N_f+k,N_f+l}\right)^2\right]\left(\delta_{mk}\delta_{nl}+\delta_{nk}\delta_{ml}\right),\label{p704}
\end{align}
where we introduced the column vector $g= ((\Gamma_m^{12})_{1,1}+1,(\Gamma_m^{12})_{2,2}+1,\dots,(\Gamma_m^{12})_{N_f,N_f}+1)^T$. Since we assume that the diagonal part of $\omega$ is zero, we set $\mathcal B_{kklm}=0=\mathcal B_{klmm}$ for all $k,l,m\in\{1,\dots,N_f\}$. The specific form of Eq.~\eqref{p704} is a direct consequence of the restrictions $\mathcal B_{klmn}=\mathcal B_{lkmn}=\mathcal B_{klnm}=\mathcal B_{lknm}$ and $\mathcal B_{klmn}=\mathcal B_{mnkl}$, which follow from the symmetry constraint $(d_\tau\omega)^T=d_\tau\omega$. By inserting Eq.~\eqref{om2} into Eq.~\eqref{om1}, we get
\begin{align}
    \frac{1}{8}\sum_{m,n=1}^{N_f}\mathcal B_{klmn}(d_\tau \omega)_{mn}=-
    \frac{dE}{d\omega_{kl}},\label{om3}
\end{align}
which can be solved for $d_\tau\omega$ by reshaping the $(N_f\times N_f\times N_f\times N_f)$-sparse tensor $\mathcal B$ into a $(N_f^2\times N_f^2)$-matrix and similarly reshape the $(N_f\times N_f)$-matrices $\omega$ and $dE/d\omega$ into column vectors of length $N_f^2$. Within that representation, we can solve for $d_\tau\omega$ by multiplying the reshaped expressions in the above equation from the left by the Moore-Penrose inverse $\mathcal B^+$ \cite{moore1920reciprocal,penrose1955generalized}.\footnote{Due to the sparseness of $\mathcal B$, its matrix representation will in general not possess full rank, which requires the pseudo-inverse.} By unfolding the pseudo-inverse $\mathcal B^+$ back into the original four-dimensional tensor form, we arrive at the following equation for the time evolution of the non-Gaussian variational parameters,
\begin{align}
    (d_\tau\omega)_{kl} = -8\sum_{m,n=1}^{N_f}\left(\mathcal B^+\right)_{klmn}\frac{dE}{d\omega_{mn}}.\label{p709}
\end{align}
We term this Ansatz the Hybrid Imaginary-Time-Gradient-Descent (HITGD) method. The main difference between the ``pure'' imaginary-time evolution of Shi \textit{et al.} that was outlined in Section~\ref{ite_shi} and the HITGD method is the replacement of the equations of motion for the non-Gaussian parameters $\omega$ at each time step $\Delta\tau$, meaning that  Eq.~\eqref{p410} is replaced by  Eq.~\eqref{p709}, while the equations of motion of the Gaussian variational parameters $\xi$ are still given by Eq.~\eqref{p105}. 

We note, that while Eq.~\eqref{p709} will always guarantee the energy to decrease monotonically, in practice it will often times be sufficient to use the simple Ansatz in Eq.~\eqref{p724}. The magnitude of the coefficient can be chosen through $c > \norm{\mathcal B}_{\text{max}}/8$, where
$\norm{\mathcal B}_{\text{max}}$ denotes the largest value of the spectral norm of the matricizised tensor $\mathcal B$.

Central to the HITGD approach is the ability to evaluate the energy gradient in Eq.~\eqref{p709}. This can be achieved by computing the commutator of the rotated Hamiltonian with the normal-ordered product of two number operators, 
\begin{align}
\frac{dE}{d\omega_{ij}}=& \frac{d}{d\omega_{ij}}\braket{\Psi_{\text{GS}}|\hat U_{\text{FA}}^\dag \hat H \hat U_{\text{FA}}|\Psi_{\text{GS}}}
=-\frac{i}{2}\braket{\Psi_{\text{GS}}|[\hat  H_{\text{FA}},\hat c_i^\dag \hat c_j^\dag \hat c_i\hat c_j]|\Psi_{\text{GS}}}.\label{p648}
\end{align}
As detailed in Appendix~\ref{en_grad}, the energy gradient in Eq.~\eqref{p648} can be computed through
\begin{align}
\frac{dE}{d\omega_{ij}}
%-----------------
=&\left(\Im\left\{\sum_{p=1}^{N_f}f_{ip}^{\text{FA}}\mathcal A^{[\alpha_{ip}^{\text{FA}}]}_{i^\dag j^\dag j p} \right\}+\Im\left\{(i\leftrightarrow j)\right\}\right)+2\Im\left\{\sum_{p<q}^{N_f}h_{ijpq}^{\text{FA}}\mathcal A^{[\beta_{ijpq}^{\text{FA}}]}_{i^\dag j^\dag pq}\right\}\nonumber\\& + 2\left(\Im\left\{\sum_{p=1}^{N_f}\sum_{q<r}h_{ipqr}^{\text{FA}}\mathcal A^{[\beta^{\text{FA}}_{ipqr}]}_{j^\dag i^\dag p^\dag j q r}\right\}+\Im\left\{(i\leftrightarrow j)\right\}\right)
,\label{p649}
\end{align}
where $\Im\{x\}=(x-x^*)/(2i)$ denotes the imaginary part of a complex number $x$ and the symbol ``$(i\leftrightarrow j)$'' is a placeholder for an expression identical to the preceding one, with the exception that the roles of the indices $i$ and $j$ are interchanged. Eq.~\eqref{p649} was derived for a general quantum many-body of the form in Eq.~\eqref{p58}, and only assumed the one- and two-body matrix element symmetries discussed in Section~\ref{qmb}.

It should be noted, that imaginary time evolution is in itself a gradient descent method with respect to the natural inner product, i.e. the Fubini-Study metric \cite{Fubini1908,Study1905,hackl2020geometry}. The terminology HITGD is used in order to distinguish the \textit{natural} (imaginary) time evolution of variational parameters described by Eq.~\eqref{p410} from the \textit{constructed} time evolution of some parameters described in this section, and in particular Eq.~\eqref{p709}. Our construction is thus only valid for a method designed to approximate the ground state of a given fermionic Hamiltonian, whereas if \textit{all} variational parameters evolve through the evaluation of Eq.~\eqref{p410}, one can not only approximate the ground state through an imaginary time evolution, but also study the dynamics of a given quantum system through a real time evolution of the system within the variational manifold \cite{SHI2018245,hackl2020geometry}. 

On a deeper level, Eq.~\eqref{p709} requires the evaluation of lower-order correlation functions compared to Eq.~\eqref{p410} hints to the fact that the gradient descent-based Ansatz employs a different metric for the non-Gaussian variational parameters than the Fubini-Study metric used for the Gaussian variational parameters. An interesting question is thus to study how the proposed HITGD method compares to the natural gradient descent (i.e. the imaginary time evolution), and if the performance can be linked to the different employed metrics.
%--------------------------------------------------------------------
%--------------------------------------------------------------------
%--------------------------------------------------------------------
%--------------------------------------------------------------------
\section{Preparing the non-Gaussian state on a quantum computer \label{qc}}
Once the classical variational method has produced a final set of parameters $\{\xi,\omega\}$, one can implement the corresponding quantum state (described by Eq.~\eqref{p2}) on a quantum computer, serving as an initial state for quantum algorithms. For near term quantum algorithms, such as the variational quantum eigensolver \cite{peruzzo2014variational}, there is only heuristic evidence indicating that a large overlap $\phi_0=\braket{\Psi_0|\Psi_{\text{initial}}}$ with the desired ground state can lower the amount of repetitions of the quantum algorithm. However, quantum algorithms relying on error-corrected quantum computers such as quantum phase estimation \cite{shor1994algorithms} or ground state preparation algorithms \cite{ge2019faster} possess a provable $\mathcal O(1/|\phi_0|^2)$, or $\mathcal O(1/|\phi_0|)$-dependence on $\phi_0$, respectively. It is therefore desirable to find initial states which are both, efficiently computable on classical computers and efficiently implementable on quantum computers, while possessing the potential to improve over an Hartree-Fock Ansatz, which is just a single fermionic Gaussian state $\ket{\Psi_{\text{GS}}}$. We presented a short sketch on how the non-Gaussian state can be realized in Fig.~\ref{fig:var_method}. The objective of this section is to detail how the black-box circuit elements for  $\hat U_{\text{FA}}(\omega)$ can be realized in terms of basic quantum logic gates.

In order to generate the fermionic Gaussian state $\ket{\Psi_{\text{GS}}}=\hat U_{\text{GS}}\ket{0}$, one can rotate the system Hamiltonian of Eq.~\eqref{p58} into the eigenbasis of the covariance matrix through orbital rotations, where the initial state is a single Slater determinant
\begin{align}
\ket{\Psi_{\text{GS}}}  = \tilde c_{k_1}^\dag \cdots \tilde c_{k_{N_f}}^\dag \ket{0},
\end{align}
where $\tilde c_{k_j}^\dag$ are the rotated fermionic operators. Alternatively, one can rotate the state into the original basis of the system Hamiltonian through Givens rotations\cite{wecker2015solving,jiang2018quantum}. The second approach is advantageous when much of the sparsity of the original system Hamiltonian would be lost and a Gaussian state preparation through $\binom{N_f}{2}$ Givens rotations would be advantageous, whereas the first approach comes at the cost of a larger number of measurements, since the number of terms in the system Hamiltonian generally increases. Construction methods to implement a fermionic Gaussian state on a quantum computer are well-known \cite{ortiz2001quantum,somma2002simulating,wecker2015solving,kivlichan2018quantum,jiang2018quantum}, have been implemented successfully on quantum hardware \cite{arute2020hartree}, and open source software is already available to transform $\hat U_{\text{GS}}$ into quantum gates \cite{mcclean2017openfermion}.

In order to realize the non-Gaussian unitary $\hat U_{\text{FA}}(\omega)$ on a quantum computer, one has to first map the fermionic operators onto the spin  operators \cite{tranter2015b}, for instance by means of the Jordan-Wigner transformation \cite{jordan1993paulische,bravyi2002fermionic}. This leads to
\begin{align}
    \hat U_{\text{FA}}(\omega)
    =& e^{\frac{i}{8}\sum_{j\neq k}\omega_{jk}} \prod_j e^{\frac{i}{4}\sum_{ k}\omega_{jk}\hat \sigma^z_j} \prod_{j<k}e^{\frac{i}{4}\omega_{jk}\hat\sigma^z_j\hat\sigma^z_k},
    \label{p403}
\end{align}
where we defined the Pauli spin-operator
\begin{align}
    \hat\sigma^z =  \begin{pmatrix}
    1&0\\0&-1
    \end{pmatrix}.\label{p404}
\end{align}
The first factor in Eq.~\eqref{p403} gives rise to a global phase and is thus irrelevant when evaluating the expectation value of the energy. The other terms lead to commuting single- and two qubit gates, with the latter being identical to Ising-type interactions, native to Josephson flux qubits  \cite{orlando1999superconducting,schuch2003natural} or inductively coupled Josephson charge qubits \cite{makhlin1999josephson}, which makes the implementation of the non-Gaussian Ansatz particularly appealing to such architectures. 

\begin{figure}
\centering
\includegraphics[width=.95\textwidth]{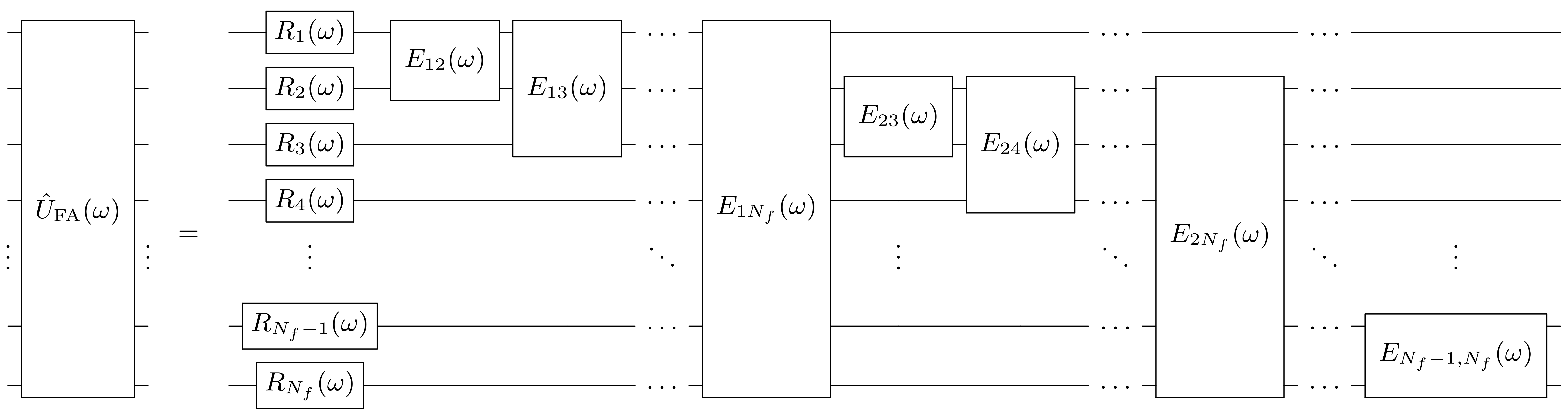}
\caption{Explicit quantum circuit which produces the non-Gaussian unitary operator $\hat U_{\text{GS}}$ using $N_f$ single-qubit $\hat \sigma^z$-rotations $R_j(\omega)=\exp\left(\frac{i}{4}\sum_k\omega_{jk}\hat\sigma^z_j\right)$ and $N_f(N_f+1)/2$ Ising-type two-qubit interactions $E_{jk}(\omega)=\exp\left(\frac{i}{4}\omega_{jk}\hat \sigma^z_j\hat\sigma^z_k\right)$. The rotation angles are determined by the final set of variational parameters $\omega$ obtained from a classical minimization of the energy functional in Eq.~\eqref{p350}. The overall phase factor present in Eq.~\eqref{p403} is omitted.\label{fig:entanglement_circuit}} 
\end{figure}

In Fig.~\ref{fig:entanglement_circuit}, we give an explicit quantum circuit that is equivalent to $\hat U_{\text{FA}}$ and requires $N_f$ single-qubit $\hat \sigma^z$-rotations and $N_f(N_f+1)/2$ two-qubit Ising-type $\hat \sigma^z_j\hat \sigma^z_k$ interactions, whose rotation angles are given by the arguments in Eq.~\eqref{p403}. The number of gates in this circuit are fixed, however, one can make use of the fact that all gates appearing in Eq.~\eqref{p403} commute with each other and follow various strategies to lower the circuit depth, which is lower bounded by $\mathcal O(\log\left(N_f\right))$. A linear SWAP-gate overhead appears, if the qubit layout does not allow for all-to-all connectivity, which is the case for most quantum architectures today. We will leave the task of optimizing the above quantum circuit to appropriate software \cite{qiskit,mcclean2017openfermion}. 

%\vspace*{2px}
%--------------------------------------------------------------------
%--------------------------------------------------------------------
%--------------------------------------------------------------------
%--------------------------------------------------------------------
\section{Summary and conclusion\label{conlclusion}}
In this work, we have presented an explicit formula to compute expectation values of fermionic operators for a non-Gaussian variational state Ansatz in Eq.~\eqref{p360}, without requiring the computation of Grassmann derivatives of a generating function. We complemented two additional expressions in Eq.~\eqref{p367}-\eqref{p368} 
which are required to compute the most general type of expression in Eq.~\eqref{p228}. In an effort to simplify the updating rule for the non-Gaussian variational parameters $\omega$, we introduced a gradient-descent-based Ansatz---inspired by the strategy of lowering the energy in each time step---in Eqs.~\eqref{p724} and \eqref{p709} which only  requires the evaluation of the energy gradient with respect to $\omega$, as opposed to the  more cumbersome state overlaps of the imaginary time-based Ansatz of Eq.~\eqref{p410}. 

While our results might benefit the simulation of strongly interacting fermionic systems on classical computers, 
the simplicity of the non-Gaussian Ansatz makes the variational method also a suitable efficient classical subroutine for providing initial states beyond Hartree-Fock states. This has the potential of lowering the run-time of various  quantum algorithms that possess an inherent dependence on the initial state overlap, such as quantum phase estimation \cite{shor1994algorithms} or sophisticated ground state preparation protocols \cite{ge2019faster}. Furthermore, the implementation of the non-Gaussian state can also serve as an efficient method to test the proficiency of large quantum computers (of fifty or more qubits) in whether they are able to reliably implement simple entangling gate structures such as those of Eq.~\eqref{p9}. 

It is important to realize that our algorithm, as well as  the Ansatz of Eq.~\eqref{p410} presented in  \cite{SHI2018245} still have to show that they can improve simulations of purely fermionic systems over a simpler generalized Hartree-Fock algorithm (which only contains a fermionic Gaussian state as its variational state). However, since the non-Gaussian Ansatz contains the family of fermionic Gaussian states, it is at least as expressive as generalized Hartree-Fock theory. 

The success of the variational method will ultimately not only be determined by the type of employed variational state Ansatz, but also by the specifics of the studied quantum system and whether physical intuition can be used to choose good initial variational parameters. We believe that future works should use the method presented in this work and apply it to explicit models, and that a proper system-dependent initial choice of the variational parameters might prove advantageous in an effort to go beyond Hartree-Fock theory. 

We note, that upon submission we became aware of a recently published work by Guaita \textit{et al.} \cite{guaita2021}, which discusses the computation of expectation values of spin, bosonic and fermionic operators with respect to an Ansatz similar to Eq.~\eqref{p2}, as well as the tangent vectors, which has a large overlap with the results presented in Section~\ref{expectation_values} and Appendix~\ref{tangent}. In order to apply the variational method to large system sizes, it is of major importance to find the numerically least expensive method for the computation of Eq.~\eqref{p228}, which is why future work should numerically compare the findings of  \cite{guaita2021} with our results. 
%--------------------------------------------------------------------
%--------------------------------------------------------------------
%--------------------------------------------------------------------
%--------------------------------------------------------------------
\section{Acknowledgement}
\texttt{OpenFermion} \cite{mcclean2017openfermion} was used to check all major analytical formulas appearing in the main text and the appendix. Figure~\ref{fig:entanglement_circuit} was created using Q-circuit \cite{eastin2004q}. The authors thank Miguel-\'{A}ngel Mart\'{i}n-Delgado and Giovanna Morigi for support and discussions. MPK thanks Murray Holland, Peter Love, Pierre-Luc Dallaire-Demers and the Zapata Computing team for hospitality and discussions. MPK acknowledges funding through S2018-TCS4342 QUITEMAD-CM. SBJ acknowledges funding from the NSF AMO Grant No. 1806827; NSF PFC Grant No. 1734006; and the DARPA and ARO Grant No. W911NF-16-1-0576. MPK and FKW acknowledge support from the EU through the FET flagship project OpenSuperQ.

\bibliography{revised_version_2021_07_15}{}

\begin{thebibliography}{10}

\bibitem{manin1980computable}
Yuri Manin.
\newblock Computable and uncomputable.
\newblock {\em Sovetskoye Radio, Moscow}, 128, 1980.

\bibitem{feynman1982simulating}
Richard~P Feynman.
\newblock Simulating physics with computers.
\newblock {\em International journal of theoretical physics}, 21(6):467--488,
  1982.

\bibitem{Lloyd1073}
Seth Lloyd.
\newblock Universal quantum simulators.
\newblock {\em Science}, 273(5278):1073--1078, 1996.

\bibitem{bernstein1997}
Ethan Bernstein and Umesh Vazirani.
\newblock Quantum complexity theory.
\newblock {\em SIAM Journal on Computing}, 26(5):1411--1473, 1997.

\bibitem{grover1996fast}
Lov~K. Grover.
\newblock A fast quantum mechanical algorithm for database search.
\newblock In {\em Proceedings of the Twenty-Eighth Annual ACM Symposium on
  Theory of Computing}, STOC '96, page 212–219, New York, NY, USA, 1996.
  Association for Computing Machinery.

\bibitem{simon1997power}
Daniel~R. Simon.
\newblock On the power of quantum computation.
\newblock In {\em Proceedings 35th Annual Symposium on Foundations of Computer
  Science}, pages 116--123, 1994.

\bibitem{shor1999polynomial}
Peter~W. Shor.
\newblock Polynomial-time algorithms for prime factorization and discrete
  logarithms on a quantum computer.
\newblock {\em SIAM J. Comput.}, 26(5):1484–1509, October 1997.

\bibitem{berry2015simulating}
Dominic~W. Berry, Andrew~M. Childs, Richard Cleve, Robin Kothari, and
  Rolando~D. Somma.
\newblock Simulating hamiltonian dynamics with a truncated taylor series.
\newblock {\em Phys. Rev. Lett.}, 114:090502, Mar 2015.

\bibitem{berry2015hamiltonian}
Dominic~W. Berry, Andrew~M. Childs, and Robin Kothari.
\newblock Hamiltonian simulation with nearly optimal dependence on all
  parameters.
\newblock In {\em 2015 IEEE 56th Annual Symposium on Foundations of Computer
  Science}, pages 792--809, 2015.

\bibitem{Low2019hamiltonian}
Guang~Hao Low and Isaac~L. Chuang.
\newblock Hamiltonian {S}imulation by {Q}ubitization.
\newblock {\em {Quantum}}, 3:163, July 2019.

\bibitem{low2017optimal}
Guang~Hao Low and Isaac~L. Chuang.
\newblock Optimal hamiltonian simulation by quantum signal processing.
\newblock {\em Phys. Rev. Lett.}, 118:010501, Jan 2017.

\bibitem{ge2019faster}
Yimin Ge, Jordi Tura, and J.~Ignacio Cirac.
\newblock Faster ground state preparation and high-precision ground energy
  estimation with fewer qubits.
\newblock {\em Journal of Mathematical Physics}, 60(2):022202, 2019.

\bibitem{kitaev1995quantum}
A.~Yu. {Kitaev}.
\newblock {Quantum measurements and the Abelian Stabilizer Problem}.
\newblock {\em arXiv e-prints}, pages quant--ph/9511026, November 1995.

\bibitem{abrams1999quantum}
Daniel~S. Abrams and Seth Lloyd.
\newblock Quantum algorithm providing exponential speed increase for finding
  eigenvalues and eigenvectors.
\newblock {\em Phys. Rev. Lett.}, 83:5162--5165, Dec 1999.

\bibitem{poulin2009preparing}
David Poulin and Pawel Wocjan.
\newblock Preparing ground states of quantum many-body systems on a quantum
  computer.
\newblock {\em Phys. Rev. Lett.}, 102:130503, Apr 2009.

\bibitem{peruzzo2014variational}
Alberto Peruzzo, Jarrod McClean, Peter Shadbolt, Man-Hong Yung, Xiao-Qi Zhou,
  {Peter J.} Love, Alan Aspuru-Guzik, and {Jeremy L.} O'Brien.
\newblock A variational eigenvalue solver on a photonic quantum processor.
\newblock {\em Nature Communications}, 5, July 2014.

\bibitem{mcclean2016theory}
Jarrod~R McClean, Jonathan Romero, Ryan Babbush, and Al{\'{a}}n Aspuru-Guzik.
\newblock The theory of variational hybrid quantum-classical algorithms.
\newblock {\em New Journal of Physics}, 18(2):023023, feb 2016.

\bibitem{Bach1993GeneralizedHT}
V.~Bach, E.~Lieb, and J.~P. Solovej.
\newblock Generalized hartree-fock theory and the hubbard model.
\newblock {\em Journal of Statistical Physics}, 76:3--89, 1993.

\bibitem{kraus2010generalized}
Christina~V Kraus and J~Ignacio Cirac.
\newblock Generalized hartree{\textendash}fock theory for interacting fermions
  in lattices: numerical methods.
\newblock {\em New Journal of Physics}, 12(11):113004, Nov 2010.

\bibitem{wick1950evaluation}
G.~C. Wick.
\newblock The evaluation of the collision matrix.
\newblock {\em Phys. Rev.}, 80:268--272, Oct 1950.

\bibitem{bravyi2005lagrangian}
Sergey Bravyi.
\newblock Lagrangian representation for fermionic linear optics.
\newblock {\em Quantum Info. Comput.}, 5(3):216–238, May 2005.

\bibitem{arute2020hartree}
Frank Arute, Kunal Arya, Ryan Babbush, Dave Bacon, Joseph~C. Bardin, Rami
  Barends, Sergio Boixo, Michael Broughton, Bob~B. Buckley, David~A. Buell,
  Brian Burkett, Nicholas Bushnell, Yu~Chen, Zijun Chen, Benjamin Chiaro,
  Roberto Collins, William Courtney, Sean Demura, Andrew Dunsworth, Edward
  Farhi, Austin Fowler, Brooks Foxen, Craig Gidney, Marissa Giustina, Rob
  Graff, Steve Habegger, Matthew~P. Harrigan, Alan Ho, Sabrina Hong, Trent
  Huang, William~J. Huggins, Lev Ioffe, Sergei~V. Isakov, Evan Jeffrey, Zhang
  Jiang, Cody Jones, Dvir Kafri, Kostyantyn Kechedzhi, Julian Kelly, Seon Kim,
  Paul~V. Klimov, Alexander Korotkov, Fedor Kostritsa, David Landhuis, Pavel
  Laptev, Mike Lindmark, Erik Lucero, Orion Martin, John~M. Martinis, Jarrod~R.
  McClean, Matt McEwen, Anthony Megrant, Xiao Mi, Masoud Mohseni, Wojciech
  Mruczkiewicz, Josh Mutus, Ofer Naaman, Matthew Neeley, Charles Neill, Hartmut
  Neven, Murphy~Yuezhen Niu, Thomas~E. O{\textquoteright}Brien, Eric Ostby,
  Andre Petukhov, Harald Putterman, Chris Quintana, Pedram Roushan, Nicholas~C.
  Rubin, Daniel Sank, Kevin~J. Satzinger, Vadim Smelyanskiy, Doug Strain,
  Kevin~J. Sung, Marco Szalay, Tyler~Y. Takeshita, Amit Vainsencher, Theodore
  White, Nathan Wiebe, Z.~Jamie Yao, Ping Yeh, and Adam Zalcman.
\newblock Hartree-fock on a superconducting qubit quantum computer.
\newblock {\em Science}, 369(6507):1084--1089, 2020.

\bibitem{kraus2009quantum}
Christina~Verena Kraus.
\newblock {\em A Quantum Information Perspective of Fermionic Quantum Many-Body
  Systems}.
\newblock Dissertation, Technische Universität München, München, 2009.

\bibitem{hackl2018aspects}
Lucas~Fabian Hackl.
\newblock Aspects of gaussian states: Entanglement, squeezing and complexity.
\newblock 2018.

\bibitem{SHI2018245}
Tao Shi, Eugene Demler, and J.~{Ignacio Cirac}.
\newblock Variational study of fermionic and bosonic systems with non-gaussian
  states: Theory and applications.
\newblock {\em Annals of Physics}, 390:245--302, 2018.

\bibitem{sala2018variational}
P.~Sala, T.~Shi, S.~K\"uhn, M.~C. Ba\~nuls, E.~Demler, and J.~I. Cirac.
\newblock Variational study of $u(1)$ and $su(2)$ lattice gauge theories with
  gaussian states in $1+1$ dimensions.
\newblock {\em Phys. Rev. D}, 98:034505, Aug 2018.

\bibitem{ashida2018solving}
Yuto Ashida, Tao Shi, Mari~Carmen Ba\~nuls, J.~Ignacio Cirac, and Eugene
  Demler.
\newblock Solving quantum impurity problems in and out of equilibrium with the
  variational approach.
\newblock {\em Phys. Rev. Lett.}, 121:026805, Jul 2018.

\bibitem{ashida2018variational}
Yuto Ashida, Tao Shi, Mari~Carmen Ba\~nuls, J.~Ignacio Cirac, and Eugene
  Demler.
\newblock Variational principle for quantum impurity systems in and out of
  equilibrium: Application to kondo problems.
\newblock {\em Phys. Rev. B}, 98:024103, Jul 2018.

\bibitem{ashida2019efficient}
Yuto Ashida, Tao Shi, Richard Schmidt, H.~R. Sadeghpour, J.~Ignacio Cirac, and
  Eugene Demler.
\newblock Efficient variational approach to dynamics of a spatially extended
  bosonic kondo model.
\newblock {\em Phys. Rev. A}, 100:043618, Oct 2019.

\bibitem{ashida2019quantum}
Yuto Ashida, Tao Shi, Richard Schmidt, H.~R. Sadeghpour, J.~Ignacio Cirac, and
  Eugene Demler.
\newblock Quantum rydberg central spin model.
\newblock {\em Phys. Rev. Lett.}, 123:183001, Oct 2019.

\bibitem{guaita2019gaussian}
Tommaso Guaita, Lucas Hackl, Tao Shi, Claudius Hubig, Eugene Demler, and
  J.~Ignacio Cirac.
\newblock Gaussian time-dependent variational principle for the bose-hubbard
  model.
\newblock {\em Phys. Rev. B}, 100:094529, Sep 2019.

\bibitem{shi2019trapped}
Tao Shi, Junqiao Pan, and Su~Yi.
\newblock Trapped bose-einstein condensates with attractive s-wave interaction.
\newblock {\em arXiv preprint arXiv:1909.02432}, 2019.

\bibitem{liu2020non}
Ruijin Liu, Yue-Ran Shi, and Wei Zhang.
\newblock Non-gaussian variational approach to fermi polarons in one- and
  two-dimensional lattices.
\newblock {\em Phys. Rev. A}, 102:033305, Sep 2020.

\bibitem{wang2020theory}
Yuqi Wang, Longfei Guo, Su~Yi, and Tao Shi.
\newblock Theory for self-bound states of dipolar bose-einstein condensates.
\newblock {\em Phys. Rev. Research}, 2:043074, Oct 2020.

\bibitem{wang2020zero}
Yao Wang, Ilya Esterlis, Tao Shi, J.~Ignacio Cirac, and Eugene Demler.
\newblock Zero-temperature phases of the two-dimensional hubbard-holstein
  model: A non-gaussian exact diagonalization study.
\newblock {\em Phys. Rev. Research}, 2:043258, Nov 2020.

\bibitem{pan2021quantum}
Junqiao Pan, Su~Yi, and Tao Shi.
\newblock Quantum phases of self-bound droplets of bose-bose mixtures.
\newblock {\em arXiv preprint arXiv:2102.02361}, 2021.

\bibitem{balian1969nonunitary}
R~Balian and E~Brezin.
\newblock Nonunitary bogoliubov transformations and extension of wick’s
  theorem.
\newblock {\em Il Nuovo Cimento B (1965-1970)}, 64(1):37--55, 1969.

\bibitem{jain2007composite}
Jainendra~K. Jain.
\newblock {\em Composite Fermions}.
\newblock Cambridge University Press, 2007.

\bibitem{cahill1999density}
Kevin~E. Cahill and Roy~J. Glauber.
\newblock Density operators for fermions.
\newblock {\em Phys. Rev. A}, 59:1538--1555, Feb 1999.

\bibitem{berezin1996method}
F.~A. Berezin.
\newblock {\em The method of second quantization, by F.A. Berezin. Translated
  by Nobumichi Mugibayashi and Alan Jeffrey}.
\newblock Academic Press New York, 1966.

\bibitem{Negele1998Quantum}
John~W. Negele and Henri Orland.
\newblock {\em Quantum Many-particle Systems}.
\newblock Westview Press, November 1998.

\bibitem{Corney2006GaussianPR}
J.~Corney and P.~Drummond.
\newblock Gaussian phase-space representations for fermions.
\newblock {\em Physical Review B}, 73:125112, 2006.

\bibitem{miller1981inverse}
Kenneth~S. Miller.
\newblock On the inverse of the sum of matrices.
\newblock {\em Mathematics Magazine}, 54(2):67--72, 1981.

\bibitem{helgaker2014molecular}
T~Helgaker, P~J{\o}rgensen, and J~Olsen.
\newblock {\em Molecular Electronic Structure Theory}.
\newblock John Wiley \& Sons, LTD, Chichester, 2000.

\bibitem{moore1920reciprocal}
Eliakim~H Moore.
\newblock On the reciprocal of the general algebraic matrix.
\newblock {\em Bull. Am. Math. Soc.}, 26:394--395, 1920.

\bibitem{penrose1955generalized}
R.~Penrose.
\newblock A generalized inverse for matrices.
\newblock {\em Mathematical Proceedings of the Cambridge Philosophical
  Society}, 51(3):406–413, 1955.

\bibitem{Fubini1908}
Guido Fubini.
\newblock Sulla teoria delle funzioni automorfe e delle loro trasformazioni.
\newblock {\em Annali di Matematica Pura ed Applicata (1898-1922)},
  14(1):33--67, Dec 1908.

\bibitem{Study1905}
E.~Study.
\newblock K{\"u}rzeste wege im komplexen gebiet.
\newblock {\em Mathematische Annalen}, 60(3):321--378, Sep 1905.

\bibitem{hackl2020geometry}
Lucas Hackl, Tommaso Guaita, Tao Shi, Jutho Haegeman, Eugene Demler, and
  Ignacio Cirac.
\newblock Geometry of variational methods: dynamics of closed quantum systems.
\newblock {\em arXiv preprint arXiv:2004.01015}, 2020.

\bibitem{shor1994algorithms}
P.~W. Shor.
\newblock Algorithms for quantum computation: discrete logarithms and
  factoring.
\newblock In {\em Proceedings 35th Annual Symposium on Foundations of Computer
  Science}, pages 124--134, 1994.

\bibitem{wecker2015solving}
Dave Wecker, Matthew~B. Hastings, Nathan Wiebe, Bryan~K. Clark, Chetan Nayak,
  and Matthias Troyer.
\newblock Solving strongly correlated electron models on a quantum computer.
\newblock {\em Phys. Rev. A}, 92:062318, Dec 2015.

\bibitem{jiang2018quantum}
Zhang Jiang, Kevin~J. Sung, Kostyantyn Kechedzhi, Vadim~N. Smelyanskiy, and
  Sergio Boixo.
\newblock Quantum algorithms to simulate many-body physics of correlated
  fermions.
\newblock {\em Phys. Rev. Applied}, 9:044036, Apr 2018.

\bibitem{ortiz2001quantum}
G.~Ortiz, J.~E. Gubernatis, E.~Knill, and R.~Laflamme.
\newblock Quantum algorithms for fermionic simulations.
\newblock {\em Phys. Rev. A}, 64:022319, Jul 2001.

\bibitem{somma2002simulating}
R.~Somma, G.~Ortiz, J.~E. Gubernatis, E.~Knill, and R.~Laflamme.
\newblock Simulating physical phenomena by quantum networks.
\newblock {\em Phys. Rev. A}, 65:042323, Apr 2002.

\bibitem{kivlichan2018quantum}
Ian~D. Kivlichan, Jarrod McClean, Nathan Wiebe, Craig Gidney, Al\'an
  Aspuru-Guzik, Garnet Kin-Lic Chan, and Ryan Babbush.
\newblock Quantum simulation of electronic structure with linear depth and
  connectivity.
\newblock {\em Phys. Rev. Lett.}, 120:110501, Mar 2018.

\bibitem{mcclean2017openfermion}
Jarrod~R McClean, Nicholas~C Rubin, Kevin~J Sung, Ian~D Kivlichan, Xavier
  Bonet-Monroig, Yudong Cao, Chengyu Dai, E~Schuyler Fried, Craig Gidney,
  Brendan Gimby, Pranav Gokhale, Thomas Häner, Tarini Hardikar, Vojt{\v{e}}ch
  Havl{\'{\i}}{\v{c}}ek, Oscar Higgott, Cupjin Huang, Josh Izaac, Zhang Jiang,
  Xinle Liu, Sam McArdle, Matthew Neeley, Thomas O'Brien, Bryan O'Gorman, Isil
  Ozfidan, Maxwell~D Radin, Jhonathan Romero, Nicolas P~D Sawaya, Bruno
  Senjean, Kanav Setia, Sukin Sim, Damian~S Steiger, Mark Steudtner, Qiming
  Sun, Wei Sun, Daochen Wang, Fang Zhang, and Ryan Babbush.
\newblock {OpenFermion}: the electronic structure package for quantum
  computers.
\newblock {\em Quantum Science and Technology}, 5(3):034014, jun 2020.

\bibitem{tranter2015b}
Andrew Tranter, Sarah Sofia, Jake Seeley, Michael Kaicher, Jarrod McClean, Ryan
  Babbush, Peter~V. Coveney, Florian Mintert, Frank Wilhelm, and Peter~J. Love.
\newblock The bravyi–kitaev transformation: Properties and applications.
\newblock {\em International Journal of Quantum Chemistry}, 115(19):1431--1441,
  2015.

\bibitem{jordan1993paulische}
P.~Jordan and E.~P. Wigner.
\newblock {\em {\"U}ber das Paulische {\"A}quivalenzverbot}, pages 109--129.
\newblock Springer Berlin Heidelberg, Berlin, Heidelberg, 1993.

\bibitem{bravyi2002fermionic}
Sergey~B. Bravyi and Alexei~Yu. Kitaev.
\newblock Fermionic quantum computation.
\newblock {\em Annals of Physics}, 298(1):210--226, 2002.

\bibitem{orlando1999superconducting}
T.~P. Orlando, J.~E. Mooij, Lin Tian, Caspar~H. van~der Wal, L.~S. Levitov,
  Seth Lloyd, and J.~J. Mazo.
\newblock Superconducting persistent-current qubit.
\newblock {\em Phys. Rev. B}, 60:15398--15413, Dec 1999.

\bibitem{schuch2003natural}
Norbert Schuch and Jens Siewert.
\newblock Natural two-qubit gate for quantum computation using the
  $\mathrm{XY}$ interaction.
\newblock {\em Phys. Rev. A}, 67:032301, Mar 2003.

\bibitem{makhlin1999josephson}
Yuriy Makhlin, Gerd Sch{\"o}n, and Alexander Shnirman.
\newblock Josephson-junction qubits with controlled couplings.
\newblock {\em Nature}, 398(6725):305, 1999.

\bibitem{qiskit}
H{\'e}ctor~Abraham et. al.
\newblock Qiskit: An open-source framework for quantum computing, 2019.

\bibitem{guaita2021}
Tommaso Guaita, Lucas Hackl, Tao Shi, Eugene Demler, and J.~Ignacio Cirac.
\newblock Generalization of group-theoretic coherent states for variational
  calculations.
\newblock {\em Phys. Rev. Research}, 3:023090, May 2021.

\bibitem{eastin2004q}
Bryan Eastin and Steven~T Flammia.
\newblock Q-circuit tutorial.
\newblock {\em arXiv preprint quant-ph/0406003}, 2004.

\bibitem{wilcox1967exponential}
R.~M. Wilcox.
\newblock Exponential operators and parameter differentiation in quantum
  physics.
\newblock {\em Journal of Mathematical Physics}, 8(4):962--982, 1967.

\bibitem{tsai2003note}
Henghsiu Tsai and K.S. Chan.
\newblock {A note on parameter differentiation of matrix exponentials, with
  applications to continuous-time modelling}.
\newblock {\em Bernoulli}, 9(5):895 -- 919, 2003.

\bibitem{najfeld1995derivatives}
I.~Najfeld and T.F. Havel.
\newblock Derivatives of the matrix exponential and their computation.
\newblock {\em Advances in Applied Mathematics}, 16(3):321--375, 1995.

\end{thebibliography}
\bibliographystyle{unsrt.bst}%
%--------------------------------------------------------------------
%--------------------------------------------------------------------
%--------------------------------------------------------------------
%--------------------------------------------------------------------
\appendix

%--------------------------------------------------------------------
%--------------------------------------------------------------------
%--------------------------------------------------------------------
%--------------------------------------------------------------------

%--------------------------------------------------------------------
%--------------------------------------------------------------------
%--------------------------------------------------------------------
%--------------------------------------------------------------------
\section{Calculation of expectation values\label{expectation}}
In this section, we will provide explicit expressions for the expectation values appearing in Eq.~\eqref{p228}, computed through the implicit formula presented in Section~\ref{part1}. Since we are computing expressions from the derivative of an exponential of the term in Eq.~\eqref{p817}, we can expand the latter as $\exp(X)=1+X+X^2/(2!)+\dots$. The argument $X=X(J_1,\dots,J_{N_f},J_1^*,\dots,J_{N_f}^*)$ is defined in Eq.~\eqref{p817} and is  a homogeneous second degree polynomial, which is evaluated at zero in the Grassmann variables $J_1,\dots,J_{N_f}$ and $J_1^*,\dots,J_{N_f}^*$ after taking the left-derivatives via the Grassmann operator defined in Eq.~\eqref{p810}. Since we are looking at a single expression, we may ignore the superscript $\tilde \alpha$ throughout this section. Also, we will here only consider the case of an identical number of creation and annihilation operators $a=b$ in the fermionic operator product for simplicity, but the result also extends to the case $a\neq b$. Since $\mathcal F_{J_{j_1}\dots J_{j_a} J_{k_1}^*\dots J_{k_a}^*}$ evaluates the Grassmann variables at zero after application of the derivative, the only non-vanishing term in the exponential expansion that survives is the $a$-th power of $X$. Therefore, we can write 
\begin{align}
\mathcal A_{j_1^\dag\dots j_a^\dag k_1\dots k_a} = s_{N_f}\left(\frac{1}{2}\right)^{N_f}\text{Pf}(\Gamma_F) \frac{1}{a!} \mathcal F_{J_{j_1}\dots J_{j_a} J_{k_1}^*\dots J_{k_a}^*} X^a.\label{p189}
\end{align}
We define the $(2N_f\times 2N_f)$-matrix 
\begin{align}
\mathcal M = \Sigma_F^\dag\text{sqrt}\left(1-e^{i\tilde\alpha}\right)\Gamma_F^{-1}\text{sqrt}\left(1-e^{i\tilde\alpha}\right)\Sigma_F,\label{p190}
\end{align}
and we rewrite Eq.~\eqref{p814} as
\begin{align}
\Gamma_F=&\text{sqrt}({1-e^{i\tilde\alpha}})\left(\Gamma_m-\frac{1+e^{i\tilde\alpha}}{1-e^{i\tilde\alpha}}\Upsilon\right) \text{sqrt}({1-e^{i\tilde\alpha}}),\label{p191}
\end{align}
where we defined the $(2N_f\times 2N_f)$-diagonal matrix 
\begin{align}
\frac{1+e^{i\tilde\alpha}}{1-e^{i\tilde\alpha}}=\mathds 1_2\otimes\begin{pmatrix}
\tfrac{1+e^{i\tilde\alpha(1)}}{1-e^{i\tilde\alpha(1)}}&&\\
&\ddots&\\
&&\tfrac{1+e^{i\tilde\alpha(N_f)}}{1-e^{i\tilde\alpha(N_f)}}
\end{pmatrix}\label{p192}.
\end{align}
We insert Eq.~\eqref{p191} into Eq.~\eqref{p190}, which leads to $\mathcal M = \Sigma_F^\dag \Gamma_H^{-1} \Sigma_F$, where we set
\begin{align}
\Gamma_H=\Gamma_m-\frac{1+e^{i\tilde\alpha}}{1-e^{i\tilde\alpha}}\Upsilon. \label{p201}
\end{align}
We use the following short-hand notation for $\Sigma_F$ and its adjoint $\Sigma_F^\dag$,
\begin{align}
\Sigma_F^\dag =&\begin{pmatrix}
\frac{-1}{1-e^{i\tilde\alpha}}&\frac{-i}{1-e^{i\tilde\alpha}}\\
\frac{e^{i\tilde\alpha}}{1-e^{i\tilde\alpha}}&\frac{-ie^{i\tilde\alpha}}{1-e^{i\tilde\alpha}}
\end{pmatrix},\
\Sigma_F =\begin{pmatrix}
\frac{e^{i\tilde\alpha}}{1-e^{i\tilde\alpha}}&\frac{-1}{1-e^{i\tilde\alpha}}\\
\frac{-ie^{i\tilde\alpha}}{1-e^{i\tilde\alpha}}&\frac{-i}{1-e^{i\tilde\alpha}}
\end{pmatrix}.\label{p203}
\end{align}
Setting $p,q\in \{1,2,\dots,N_f\}$, we first compute $\mathcal M_{q,p}$, which is restricted to the upper-left block of $\mathcal M$,
\begin{align}
\mathcal M_{q,p} =& \left(\mathbf 1_q^T,\mathbf 0^T\right)\Sigma_F^\dag\Gamma_H^{-1}\Sigma_F \begin{pmatrix}
\mathbf 1_p\\
\mathbf 0
\end{pmatrix}%= \mathbf 1_q^T\begin{pmatrix}
%\frac{-1}{1-e^{i\tilde\alpha}},\frac{-i}{1-e^{i\tilde\alpha}}
%\end{pmatrix}\Gamma_H^{-1}\begin{pmatrix}
%\frac{e^{i\tilde\alpha}}{1-e^{i\tilde\alpha}}\\
%\frac{-ie^{i\tilde\alpha}}{1-e^{i\tilde\alpha}}
%\end{pmatrix} 
%\mathbf 1_p\nonumber\\
=\frac{-1}{1-e^{i\tilde\alpha(q)}}\frac{e^{i\tilde\alpha(p)}}{1-e^{i\tilde\alpha(p)}}\begin{pmatrix}
\mathbf 1_q^T,\mathbf i_q^T
\end{pmatrix}\Gamma_H^{-1}\begin{pmatrix}
\mathbf 1_p\\
-\mathbf i_p
\end{pmatrix}.\label{p204}
\end{align}
In an analogous manner, we compute the matrix element $\mathcal M_{p+N_f,q+N_f}$ of the lower-right block of $\mathcal M$, giving 
\begin{align}
\mathcal M_{p+N_f,q+N_f}%=& \left(\mathbf 0^T,\mathbf 1_p^T\right)\Sigma_F^\dag\Gamma_H^{-1}\Sigma_F \begin{pmatrix}
%\mathbf 0\\
%\mathbf 1_q
%\end{pmatrix}
= \frac{e^{i\tilde\alpha(p)}}{1-e^{i\tilde\alpha(p)}}\frac{-1}{1-e^{i\tilde\alpha(q)}}\begin{pmatrix}
\mathbf 1_p^T,-\mathbf i_p^T
\end{pmatrix}\Gamma_H^{-1}\begin{pmatrix}
\mathbf 1_q\\
\mathbf i_q
\end{pmatrix}. \label{p205}
\end{align}
Since $\Gamma_H$ is skew-symmetric, it follows that $\mathcal M_{p+N_f,q+N_f}=-\mathcal M_{q,p}$. We now turn to the cross terms $\mathcal M_{q, p+N_f}$ and $\mathcal M_{p+N, q}$. The upper-right matrix elements of the block matrix $\mathcal M$ are given by
\begin{align}
\mathcal M_{q,p+N_f} 
=&\frac{1}{\left(1-e^{i\tilde\alpha(q)}\right)\left(1-e^{i\tilde\alpha(p)}\right)}\begin{pmatrix}
\mathbf 1_q^T,\mathbf i_q^T
\end{pmatrix}\Gamma_H^{-1}\begin{pmatrix}
\mathbf 1_p\\
\mathbf i_p
\end{pmatrix}=-\mathcal M_{p,q+N_f}, \label{p207}
\end{align}
while the elements of the lower-left block matrix are
\begin{align}
\mathcal M_{p+N_f,q} 
=&\frac{1}{\left(1-e^{-i\tilde\alpha(p)}\right)\left(1-e^{-i\tilde\alpha(q)}\right)}\begin{pmatrix}
\mathbf 1_p^T,-\mathbf i_p^T
\end{pmatrix}\Gamma_H^{-1}\begin{pmatrix}
\mathbf 1_q\\
-\mathbf i_q
\end{pmatrix}=-\mathcal M_{q+N_f,p}\label{p208}.
\end{align}
We now consider the zeroth-order monomial expression, which is simply given by $\mathcal A
=s_{N_f}2^{-N_f} \text{Pf}(\Gamma_F)$. This result is identical to Eq.~(D.10) in  \cite{SHI2018245}. Turning to expectation values of a product of two fermionic operators with respect to the non-Gaussian state, we compute $\mathcal A_{p^\dag q} = \mathcal A\mathcal F_{J_p J_q^*}X$, where
\begin{align}
X
=&-\frac{i}{2}\sum_{v,w=1}^{N_f}\left(J_v^*\mathcal M_{v,w}J_w+J_v^*\mathcal M_{v,w+N_f}J_w^*+J_v\mathcal M_{v+N_f,w}J_w+J_v\mathcal M_{v{+N_f},w+N_f}J_w^*\right)+\sum_{k=1}^{N_f} \frac{J_k^*J_k}{1-e^{-i\tilde\alpha(k)}}\label{p199}
\end{align}
is the argument of the exponential and identical to Eq.~\eqref{p817}. We need to reorder the Grassmann variables in order to apply the left-derivative  \cite{cahill1999density}. The first derivative is given by $\delta/(\delta J_q^*)$, therefore all conjugate Grassmann variables need to be moved to the left. Once the term in Eq.~\eqref{p199} is reordered, we can take the derivative operator defined in Eq.~\eqref{p810}, resulting in 
\begin{align}
\mathcal F_{J_p J_q^*}X
=& 
\left( \tfrac{\delta_{pq}}{1-e^{-i\tilde\alpha(q)}}-i\mathcal M_{q,p}\right).\label{p200}
\end{align}
Plugging Eq.~\eqref{p204} into Eq.~\eqref{p200} leads to
\begin{align}
\mathcal A_{p^\dag q}
=&\mathcal A\left( \frac{\delta_{pq}}{1-e^{-i\tilde\alpha(q)}}+\frac{ie^{i\tilde\alpha(p)}}{\left(1-e^{i\tilde\alpha(p)}\right)\left(1-e^{i\tilde\alpha(q)}\right)}\begin{pmatrix}
\mathbf 1_q^T,\mathbf i_q^T
\end{pmatrix}\Gamma_H^{-1}\begin{pmatrix}
\mathbf 1_p\\
-\mathbf i_p
\end{pmatrix}\right)\label{p209}.
\end{align}
The above result should coincide with Eq.~(D.11) in  \cite{SHI2018245}, which states that 
\begin{align}
\mathcal A_{p^\dag q}=&\mathcal A\frac{i}{4}e^{i\tilde\alpha_p}\begin{pmatrix}
\mathbf 1_q^T,\mathbf i_q^T
\end{pmatrix}\mathcal G\begin{pmatrix}
\mathbf 1_p\\
-\mathbf i_p
\end{pmatrix},\label{p210}
\end{align}
where $\mathcal G$ is identical to Eq.~\eqref{p365}. We now show that the expressions in Eq.~\eqref{p209} and Eq.~\eqref{p210} are equivalent. To achieve this, we first rewrite the denominator of on the right-hand side of Eq.~\eqref{p365} as 
\begin{align}
\mathds 1_{2N_f}+\frac{1}{2}\left(1-e^{i\tilde\alpha}\right)\left(\Upsilon\Gamma_m-\mathds 1_{2N_f}\right)=&\frac{1}{2}\left(1-e^{i\tilde\alpha}\right)\Upsilon\Gamma_H,\label{p211}
\end{align}
where we used that $\Upsilon^2=-\mathds 1_{2N_f}$ and that the commutator of $\Upsilon$ and $(1+e^{i\tilde\alpha})(1-e^{i\tilde\alpha})^{-1}$ is zero. We can then write
\begin{align}
\mathcal G
=&2(\Gamma_m+\Upsilon)\left(\left(1-e^{i\tilde\alpha}\right)\Upsilon\Gamma_H\right)^{-1}=2\left(-\Upsilon\frac{1}{1-e^{i\tilde\alpha}}+\frac{2}{1-e^{i\tilde\alpha}}\Upsilon\Gamma_H^{-1}\Upsilon^{-1}\frac{1}{1-e^{i\tilde\alpha}}\right).\label{p212}
\end{align}
Note, that for any $\tilde\alpha$, the matrix $\mathcal G^{[\tilde\alpha]}$ is skew-symmetric, since
\begin{align}
\left(\mathcal G^{[\tilde\alpha]}\right)^T=&- \left[\mathds 1_{2N_f}+\frac{1}{2}\left(\Gamma_m\Upsilon-\mathds 1_{2N_f}\right)\left(1-e^{i\tilde\alpha}\right)\right]^{-1}(\Gamma_m+\Upsilon)\nonumber\\
%%%
=&-(\Gamma_m+\Upsilon)\left[\mathds 1_{2N_f}+\tfrac{1}{2}\left(1-e^{i\tilde\alpha}\right)\left(\Upsilon\Gamma_m-\mathds 1_{2N_f}\right)\right]^{-1}
=-\mathcal G^{[\tilde\alpha]}.\label{p593}
\end{align}
We insert Eq.\eqref{p212} into Eq.~\eqref{p210} and note that $(\mathbf 1_q^T,\mathbf i_k^T)$ and $(\mathbf 1_q,\mathbf i_k)^T$ are left- and right-eigenvectors of $\Upsilon$ and therefore also of its inverse, resulting in 
\begin{align}
\mathcal A_{p^\dag q}%=&\mathcal A\frac{i}{2}e^{i\tilde\alpha(p)}\begin{pmatrix}
%\mathbf 1_q^T,\mathbf i_q^T
%\end{pmatrix}\left(-\Upsilon\frac{1}{1-e^{i\tilde\alpha}}+\frac{2}{1-e^{i\tilde\alpha}}\Upsilon\Gamma_H^{-1}\Upsilon^{-1}\frac{1}{1-e^{i\tilde\alpha}}\right)\begin{pmatrix}
%\mathbf 1_p\\
%-\mathbf i_p
%\end{pmatrix}\nonumber\\
=&\mathcal A\left(\frac{\delta_{p q}}{1-e^{-i\tilde\alpha(q)}}+\frac{ie^{i\tilde\alpha(p)}}{\left(1-e^{i\tilde\alpha(p)}\right)\left(1-e^{i\tilde\alpha(q)}\right)}\begin{pmatrix}
\mathbf 1_q^T,\mathbf i_q^T
\end{pmatrix}\Gamma_H^{-1}\begin{pmatrix}
\mathbf 1_p\\
-\mathbf i_p
\end{pmatrix}\right),\label{p213}
\end{align}
which is identical to  Eq.~\eqref{p209}, and therefore our result is in exact agreement with Eq.~(D.11) in  \cite{SHI2018245}. An alternative expression for $\mathcal A_{p^\dag q}$ in terms of the matrix elements $\mathcal M$ is given by 
\begin{align}
\mathcal A_{p^\dag q}=&\mathcal A\left( \frac{\delta_{pq}}{1-e^{-i\tilde\alpha(q)}}-i\mathcal M_{q,p}\right)\label{p222}.
\end{align}
The representation in Eq.~\eqref{p222} in terms of $\mathcal M$ will be useful for finding underlying symmetries, while the representation in terms of $\mathcal G$ in Eq.~\eqref{p210} gives a desirable form for numerical implementation, as it avoids numerical pitfalls which can appear for values $\tilde\alpha(j)=0$. We now turn to the cross terms $\mathcal A_{p^\dag q^\dag}$ and $\mathcal A_{pq}$, which are not considered in  \cite{SHI2018245}. Using Eq.~\eqref{p189}, we compute
\begin{align}
\lim\limits_{J\rightarrow 0}\frac{d}{d J_p}\frac{d}{d J_q} X
=&i\mathcal M_{p+N_f,q},\label{p297}\\
\lim\limits_{J\rightarrow 0}\frac{d}{d J_p^*}\frac{d}{d J_q^*}
X
=&i\mathcal M_{p,q+N_f}.\label{p298}
\end{align}
This leads to
\begin{align}
  \mathcal A_{p^\dag q^\dag}=&-i\mathcal A\mathcal M_{q+N_f,p}
  =
\frac{-i\mathcal A}{\left(1-e^{-i\tilde\alpha(p)}\right)\left(1-e^{-i\tilde\alpha(q)}\right)}\begin{pmatrix}
\mathbf 1_q^T,-\mathbf i_q^T
\end{pmatrix}\Gamma_H^{-1}\begin{pmatrix}
\mathbf 1_p\\
-\mathbf i_p
\end{pmatrix}  \label{p299}\\
\mathcal A_{p q}=&-i\mathcal A\mathcal M_{q,p+N_f}
= \frac{-i\mathcal A}{\left(1-e^{i\tilde\alpha(q)}\right)\left(1-e^{i\tilde\alpha(p)}\right)}\begin{pmatrix}
\mathbf 1_q^T,\mathbf i_q^T
\end{pmatrix}\Gamma_H^{-1}\begin{pmatrix}
\mathbf 1_p\\
\mathbf i_p
\end{pmatrix}\label{p300}.
\end{align}
We first turn to Eq.~\eqref{p299}. Using Eq.~\eqref{p208} and Eq.~\eqref{p212}, we compute
\begin{align}
    \begin{pmatrix}
\mathbf 1_q^T,-\mathbf i_q^T
\end{pmatrix}\mathcal G\begin{pmatrix}
\mathbf 1_p\\
-\mathbf i_p
\end{pmatrix}
=&\frac{-4e^{-i(\tilde\alpha(p)+\tilde\alpha(q))}}{\left(1-e^{-i\tilde\alpha(q)}\right)\left(1-e^{-i\tilde\alpha(p)}\right)} \begin{pmatrix}
\mathbf 1_q^T,-\mathbf i_q^T
\end{pmatrix}\Gamma_H^{-1}\begin{pmatrix}
\mathbf 1_p\\
-\mathbf i_p
\end{pmatrix}.  \label{p301}
\end{align}
By comparing the result of Eq~\eqref{p301} with Eq.~\eqref{p299}, we arrive at the following expression,
\begin{align}
\mathcal A_{p^\dag q^\dag}=&\frac{i}{4}\mathcal Ae^{i(\tilde\alpha(p)+\tilde\alpha(q))}\begin{pmatrix}
\mathbf 1_q^T,-\mathbf i_q^T
\end{pmatrix}\mathcal G\begin{pmatrix}
\mathbf 1_p\\
-\mathbf i_p
\end{pmatrix}.\label{p305}
\end{align}
Analogously, we compute
\begin{align}
    \begin{pmatrix}
\mathbf 1_q^T,\mathbf i_q^T
\end{pmatrix}\mathcal G\begin{pmatrix}
\mathbf 1_p\\
\mathbf i_p
\end{pmatrix}
=&\frac{-4}{\left(1-e^{-i\tilde\alpha(p)}\right)\left(1-e^{-i\tilde\alpha(q)}\right)} \begin{pmatrix}
\mathbf 1_q^T,\mathbf i_q^T
\end{pmatrix}\Gamma_H^{-1}\begin{pmatrix}
\mathbf 1_p\\
\mathbf i_p
\end{pmatrix},  \label{p304}
\end{align}
which leads to 
\begin{align}
\mathcal A_{p q}=&\frac{i}{4}\mathcal A\begin{pmatrix}
\mathbf 1_q^T,\mathbf i_q^T
\end{pmatrix}\mathcal G\begin{pmatrix}
\mathbf 1_p\\
\mathbf i_p
\end{pmatrix}.  \label{p306}
\end{align}
The expressions in Eqs.~\eqref{p305} and \eqref{p306} are essential for the evaluation of the general expectation value in Eq.~\eqref{p228} through Eq.~\eqref{p360}.

We can use the same strategy to compute non-Gaussian expectation values of quartic fermionic monomials, such as
\begin{align}
\mathcal A_{p^\dag q^\dag rs}
=&\mathcal A \frac{1}{2!} \mathcal F_{J_p J_q J_r^*J_s^*} X^2,\label{p214}
\end{align}
by first computing $X^2$, where $X$ is given by Eq.~\eqref{p199}. Since the derivative operator will only give non-zero contributions to terms where the number $a$ ($=2$) of non-conjugated Grassmann variables equals the number $a$ ($=2$) of conjugated variables, we can drop all other terms. The derivatives are ordered in such a way that the conjugated Grassmann variables are acted on first (i.e. on the left). Therefore, we reorder the remaining terms, simplifying the relevant part of the squared expression $X^2$ to get
\begin{align}
&\sum_{k,l=1}^{N_f} \frac{-J_k^*J_l^*J_kJ_l}{\left(1-e^{-i\tilde\alpha(k)}\right)\left(1-e^{-i\tilde\alpha(l)}\right)}+i\sum_{k,v,w=1}^{N_f} \frac{J_k^*J_v^*J_k J_w\mathcal M_{v,w}-J_k^*J_w^*J_kJ_v\mathcal M_{v{+N_f},w+N_f}}{1-e^{-i\tilde\alpha(k)}}\nonumber\\&+\sum_{k,l,v,w=1}^{N_f}J_k^*J_v^*J_lJ_w\mathcal M_{k,l}\mathcal M_{v,w}-\frac{1}{2}\sum_{k,l,v,w=1}^{N_f} J_k^*J_l^*J_vJ_w\mathcal M_{k,l+N_f}\mathcal M_{v+N_f,w},\label{p215}
\end{align}
where we again stress that the above terms are the only non-vanishing terms of $X^2$ under the action of the derivative operator for a quartic fermionic monomial. 
We can use $\frac{\delta}{\delta \alpha_n} \alpha_j\alpha_k= \delta_{nj}\alpha_k - \delta_{nk}\alpha_j$ to derive the following relations,
\begin{align}
\frac{\delta^2(\alpha_j\alpha_k)}{\delta \alpha_m\delta \alpha_n} =& \delta_{nj}\delta_{mk} - \delta_{nk}\delta_{mj},\label{p217}\\
\frac{\delta^4(\alpha_i^*\alpha_j^*\alpha_k\alpha_l)}{\delta \alpha_p\delta \alpha_q\delta \alpha_r^*\delta \alpha_s^*} 
=&\delta_{si}\delta_{rj}\delta_{qk}\delta_{pl}-\delta_{si}\delta_{rj}\delta_{ql}\delta_{pk}-\delta_{sj}\delta_{ri}\delta_{qk}\delta_{pl}+\delta_{sj}\delta_{ri}\delta_{ql}\delta_{pk}.\label{p218}
\end{align}
Applying $\mathcal F_{J_p J_q J_r^*J_s^*}$ on $X^2$, the only non-vanishing part that needs to be considered is displayed in Eq.~\eqref{p215}. We will list the results for each of the four sums below. The derivative contributions from the first sum result in
\begin{align}
\mathcal F_{J_p J_q J_r^*J_s^*}\sum_{k,l=1}^{N_f} \frac{-J_k^*J_l^*J_kJ_l}{\left(1-e^{-i\tilde\alpha(k)}\right)\left(1-e^{-i\tilde\alpha(l)}\right)}
=& \frac{2\left(\delta_{ps}\delta_{qr}-\delta_{qs}\delta_{pr}\right)}{\left(1-e^{-i\tilde\alpha(q)}\right)\left(1-e^{-i\tilde\alpha(p)}\right)}\label{p219},
\end{align}
the derivative on the second sum in Eq.~\eqref{p215} gives
\begin{align}
i\mathcal F_{J_p J_q J_r^*J_s^*}\sum_{k,v,w=1}^{N_f}\frac{J_k^*J_v^*J_k J_w\mathcal M_{v,w}-J_k^*J_w^*J_kJ_v\mathcal M_{v{+N_f},w+N_f}}{1-e^{-i\tilde\alpha(k)}}=&\frac{2i\left(\delta_{qs}\mathcal M_{r,p}-\delta_{qr}\mathcal M_{s,p}\right)}{1-e^{-i\tilde\alpha(q)}}-\frac{2i\left(\delta_{ps}\mathcal M_{r,q}-\delta_{pr}\mathcal M_{s,q}\right)}{1-e^{-i\tilde\alpha(p)}}.\label{p220}
\end{align}
Applying the derivative operator on the third and fourth terms of Eq.~\eqref{p215}, we end up with the following expression,
\begin{align}
&\mathcal F_{J_p J_q J_r^*J_s^*}\sum_{k,l,v,w=1}^{N_f}J_k^*J_v^*J_lJ_w\mathcal M_{k,l}\mathcal M_{v,w}-\frac{1}{2}\mathcal F_{J_p J_q J_r^*J_s^*}\sum_{k,l,v,w=1}^{N_f} J_k^*J_l^*J_vJ_w\mathcal M_{k,l+N_f}\mathcal M_{v+N_f,w}\nonumber\\
=&2\left(\mathcal M_{r,p}\mathcal M_{s,q}-\mathcal M_{r,q}\mathcal M_{s,p}-\mathcal M_{s,r+N_f}\mathcal M_{q+N_f,p}\right).\label{p221}
\end{align}
Inserting the results of Eqs.~\eqref{p219}-\eqref{p221} into Eq.~\eqref{p214}, we get 
\begin{align}
\mathcal A_{p^\dag q^\dag rs} %=& \mathcal A\left(\frac{\left(\delta_{ps}\delta_{qr}-\delta_{qs}\delta_{pr}\right)}{\left(1-e^{-i\tilde\alpha(q)}\right)\left(1-e^{-i\tilde\alpha(p)}\right)}+\frac{i\delta_{qs}}{1-e^{-i\tilde\alpha(q)}}\mathcal M_{r,p}\right.\nonumber\\&\left.-\frac{i\delta_{ps}}{1-e^{-i\tilde\alpha(p)}}\mathcal M_{r,q}-\frac{i\delta_{qr}}{1-e^{-i\tilde\alpha(q)}}\mathcal M_{s,p}\right.\nonumber\\&\left.+\frac{i\delta_{pr}}{1-e^{-i\tilde\alpha(p)}}\mathcal M_{s,q}+\mathcal M_{r,p}\mathcal M_{s,q}-\mathcal M_{r,q}\mathcal M_{s,p}-\mathcal M_{s,r+N_f}\mathcal M_{q+N_f,p} \right)\nonumber\\
=& \mathcal A\left(\left(\frac{\delta_{ps}}{1-e^{-i\tilde\alpha(p)}}-i\mathcal M_{s,p}\right)\left(\frac{\delta_{qr}}{1-e^{-i\tilde\alpha(q)}}-i\mathcal M_{r,q}\right)\right.\nonumber\\
&\left.-\left(\frac{\delta_{pr}}{1-e^{-i\tilde\alpha(p)}}-i\mathcal M_{r,p}\right)\left(\frac{\delta_{qs}}{1-e^{-i\tilde\alpha(q)}}-i\mathcal M_{s,q}\right)+(-i)^2\mathcal M_{q+N_f,p}\mathcal M_{s,r+N_f}\right).\label{p223}
\end{align}
Eq.~\eqref{p223} was written in order to highlight the factorization of the fourth-order expression into second-order expressions, identical to the splitting in the second line of Eq.~\eqref{p223} with respect to the the second-order monomials as given in Eq.~\eqref{p222}, i.e. 
\begin{align}
    \frac{\mathcal A_{p^\dag q^\dag rs}}{\mathcal A}=&\left( \frac{\mathcal A_{p^\dag s}}{\mathcal A}\frac{\mathcal A_{q^\dag r}}{\mathcal A} - \frac{\mathcal A_{p^\dag r}}{\mathcal A}\frac{\mathcal A_{q^\dag s}}{\mathcal A} + \frac{\mathcal A_{p^\dag q^\dag}}{\mathcal A}\frac{\mathcal A_{rs}}{\mathcal A} \right)\label{p357}.
\end{align}
These computations can be extended analogously to higher-order terms. 
%--------------------------------------------------------------------
%--------------------------------------------------------------------
%--------------------------------------------------------------------
%--------------------------------------------------------------------
\section{Mean-field matrices for the variational method\label{mf_var_method}}
In this appendix, we will give explicit formulas for the mean-field terms ${H_{\text{FA}}}_m$ and $O_m$ in Section~\ref{H_m} and Section~\ref{O_m}, respectively. While the expression for $O_m$ is general, the derived expression for ${H_{\text{FA}}}_m$ depends on the form of the system Hamiltonian and is here derived for systems described by the quantum many-body Hamiltonian of Eq.~\eqref{p58}. The derived expressions are crucial for the computation of the evolution of the variational parameters  in Eq.~\eqref{p105}. 
%--------------------------------------------------------------------
\subsection{Computing the mean-field energy matrix\label{H_m}}
Central to the computation of the mean-field matrix will be the computation of the derivative of expectation values with respect to the covariance (and thus with respect to a structured-) matrix. We compute 
\begin{align}
\frac{d\mathcal G^{[\tilde\alpha]}}{d\left(\Gamma_m\right)_{ij}}=&\frac{1}{2}\mathcal L^{[\tilde\alpha]}\left(\ket{i}\bra{j}-\ket{j}\bra{i}\right)\left(\mathcal L^{[\tilde\alpha]}\right)^T,\label{p521}\\
\frac{d\mathcal A^{[\tilde\alpha]}}{d (\Gamma_m)_{ij}}=&\mathcal Q^{[\tilde\alpha]}_{ij}\mathcal A^{[\tilde\alpha]},\label{p527}
\end{align}
where we defined the $(2N_f\times 2N_f)$-matrices
\begin{align}
\mathcal L^{[\tilde\alpha]} =&\mathds 1_{2N_f}-\frac{1}{2}\mathcal G^{[\tilde\alpha]}\left(1-e^{i\tilde\alpha}\right)\Upsilon,\label{p485}\\
\mathcal Q^{[\tilde\alpha]}=& -\frac{1}{2}\text{sqrt}\left(1-e^{i\tilde \alpha}\right){\left(\Gamma_F^{[\tilde\alpha]}\right)}^{-1}\text{sqrt}\left(1-e^{i\tilde \alpha}\right)\label{p487}.
\end{align}
Note that the form of e.g. Eq.~\eqref{p521} is a result of taking the derivative with respect to a structured (here, skew-symmetric) matrix. With Eqs.~\eqref{p521}-\eqref{p527}, we can compute the mean-field terms for the one- and two-body terms of the Hamiltonian of Eq.~\eqref{p58}, $\left({H_{\text{FA}}}_m\right)_{ij}=\left({H_{\text{FA},1}}_{m}\right)_{ij}+\left({H_{\text{FA},2}}_{m}\right)_{ij}$, where 
\begin{align}
\left({H_{\text{FA},1}}_{m}\right)_{ij} =&i\sum_{p,q=1}^{N_f}f_{pq}\mathcal A^{[\alpha_{pq}^{\text{FA}}]}\left[\mathcal Q^{[\alpha_{pq}^{\text{FA}}]}_{ij}\left(\mathcal G^{[\alpha_{pq}^{\text{FA}}]}\right)_{pq}^{+-}\right.\nonumber\\&\left.+\frac{1}{2}\left(\mathcal L^{[\alpha_{pq}^{\text{FA}}]}\left(\ket{i}\bra{j}-\ket{j}\bra{i}\right)\left(\mathcal L^{[\alpha_{pq}^{\text{FA}}]}\right)^T\right)_{pq}^{+-}\right],\label{p531}\\
%%%%
\left({H_{\text{FA},2}}_{m}\right)_{ij}=&-\frac{1}{16}\sum_{p,q,r,s=1}^{N_f}h_{pqrs}e^{i(\omega_{rs}-\omega_{pq})}\mathcal A^{[\beta_{pqrs}^{\text{FA}}]}\left[4\mathcal Q^{[\beta_{pqrs}^{\text{FA}}]}_{ij}\left(\mathcal G^{[\beta_{pqrs}^{\text{FA}}]}\right)_{ps}^{+-}\left(\mathcal G^{[\beta_{pqrs}^{\text{FA}}]}\right)_{qr}^{+-}
\right.\nonumber\\
&\left.+2\mathcal Q^{[\beta_{pqrs}^{\text{FA}}]}_{ij}\left(\mathcal G^{[\beta_{pqrs}^{\text{FA}}]}\right)_{pq}^{++}
\left(\mathcal G^{[\beta_{pqrs}^{\text{FA}}]}\right)_{rs}^{--}\right.\nonumber\\&\left.+4\left(\mathcal L^{[\beta_{pqrs}^{\text{FA}}]}(\ket{i}\bra{j}-\ket{j}\bra{i})\left(\mathcal L^{[\beta_{pqrs}^{\text{FA}}]}\right)^T\right)_{ps}^{+-}\left(\mathcal G^{[\beta_{pqrs}^{\text{FA}}]}\right)_{qr}^{+-}\right.\nonumber\\&\left.+\left(\mathcal L^{[\beta_{pqrs}^{\text{FA}}]}(\ket{i}\bra{j}-\ket{j}\bra{i})\left(\mathcal L^{[\beta_{pqrs}^{\text{FA}}]}\right)^T\right)_{pq}^{++}\left(\mathcal G^{[\beta_{pqrs}^{\text{FA}}]}\right)_{rs}^{--}\right.\nonumber\\&\left.+\left(\mathcal G^{[\beta_{pqrs}^{\text{FA}}]}\right)_{pq}^{++}\left(\mathcal L^{[\beta_{pqrs}^{\text{FA}}]}(\ket{i}\bra{j}-\ket{j}\bra{i})\left(\mathcal L^{[\beta_{pqrs}^{\text{FA}}]}\right)^T\right)_{rs}^{--}\right].\label{p532}
\end{align}
By using relations of the form
\begin{align}
&\left(\mathcal L^{[\tilde \alpha]}\left(\ket{i}\bra{j}-\ket{j}\bra{i}\right)\left(\mathcal L^{[\tilde \alpha]}\right)^T\right)_{ps}^{+-}\nonumber\\
=&\bra{i}\left(\mathcal L^{[\tilde \alpha]}\right)^T\left(\begin{pmatrix}
\mathbf 1_s\\
\mathbf i_s
\end{pmatrix}
\begin{pmatrix}
\mathbf 1_p^T,-\mathbf i_p^T
\end{pmatrix}-\begin{pmatrix}
\mathbf 1_p\\
-\mathbf i_p
\end{pmatrix}\begin{pmatrix}
\mathbf 1_s^T,\mathbf i_s^T
\end{pmatrix}\right)\mathcal L^{[\tilde \alpha]}\ket{j},\label{p565}
\end{align}
one can recast Eqs.~\eqref{p531}-\eqref{p532} into a matrix equation, which is well-suited for numerical implementation. 

In the rest of this subsection, we discuss a strategy to speed up the numerical evaluation of Eq.~\eqref{p487}. We can apply the algorithm of  \cite{miller1981inverse} in order to speed up the computation of $\mathcal Q^{[\tilde \alpha]}$ as defined in Eq.~\eqref{p487},
\begin{align*}
\mathcal Q^{[\tilde\alpha]}=& -\frac{1}{2}\text{sqrt}\left(1-e^{i\tilde \alpha}\right){\left(\Gamma_F^{[\tilde\alpha]}\right)}^{-1}\text{sqrt}\left(1-e^{i\tilde \alpha}\right),
\end{align*}
as it contains the inverse of the altered covariance matrix $\Gamma_F^{[\tilde\alpha]}$ as defined in Eq.~\eqref{p814}. We rewrite Eq.~\eqref{p487} as 
\begin{align}
\mathcal Q^{[\tilde\alpha]}=&-\frac{1}{2}\left(\Gamma_m-\sigma\otimes \text{diag}\left(\frac{1+e^{i\tilde\alpha}}{1-e^{i\tilde\alpha}}\right)\right)^{-1}, \label{p1036}
\end{align}
and identify $A =\Gamma_m$ and 
\begin{align}
    B =& -\sigma\otimes \text{diag}\left(\frac{1+e^{i\tilde\alpha}}{1-e^{i\tilde\alpha}}\right).\label{p1032}
\end{align}
For pure fermionic Gaussian states $A^{-1} = -\Gamma_m$. We now write $B$ as a sum of rank 1 matrices $B_k$. We again introduce a list $[i_1,\dots,i_{r_B}]$ that contains all indices whose corresponding coefficient $\tilde\beta_k=(1+e^{i\tilde\alpha_{i_k}})/(1-e^{i\tilde\alpha_{i_k}})$ is non-zero. By symmetry, $r_B$ is an even number and we have $\tilde\beta_k=\tilde\beta_{k+r_B/2} \ \forall k\leq r_B/2$. A natural choice for the decomposition of $B$ into rank 1 matrices is thus 
\begin{align}
    B= -\sum_{k\leq \frac{r_B}{2}}\tilde\beta_k\ket{i_k}\bra{i_{k}+N_f}+\sum_{k> \frac{r_B}{2}}\tilde\beta_k\ket{i_k}\bra{i_{k}-N_f}.\label{p1034}
\end{align}
We can write Eq.~\eqref{p1036} as $\mathcal Q^{[\tilde\alpha]}=-\frac{1}{2}[A+B]^{-1}$, where the inverse can be computed through the method described in \cite{miller1981inverse}.

Note, that the results of  \cite{miller1981inverse} can only be applied here when $e^{i\tilde\alpha(j)}\neq 1 \ \forall j =1,\dots,N_f$. As soon as at least one of the elements of $\tilde\alpha(j)=2m\pi$ with $m\in \mathds Z$, one has to compute $\mathcal Q^{[\tilde \alpha]}$ using Eq.~\eqref{p487} and thus through a regular computation of the matrix inverse of the expression $\Gamma_F^{[\tilde\alpha]}$. For the particular case, that all $\tilde\alpha_j=2m \pi$ with $m\in \mathds Z$, we have $\mathcal Q^{[\tilde\alpha = 2m\pi]}=\mathbb 0_{2N_f}$. 
%--------------------------------------------------------------------
\subsection{Computing the non-Gaussian mean-field matrix\label{O_m}}
The explicit expression of ${O}_m$ can be obtained from
\begin{align}
    \left(O_{m}\right)_{ij} 
    =& -2i\sum_{k,l=1}^{N_f} \left(d_\tau  \omega_{kl}\right)\frac{d\braket{\Psi_{\text{GS}}|\hat c_k^\dag \hat c_l^\dag \hat c_k\hat c_l|\Psi_{\text{GS}}}}{d\left(\Gamma_m\right)_{ij}}\label{p435}.
\end{align}
For the above case, Eq.~\eqref{p521} simplifies to  $d\mathcal G^{[0]}/(d\left(\Gamma_m\right)_{ij})=\frac{1}{2}(\ket{i}\bra{j}-\ket{j}\bra{i})$, which results in 
\begin{align}
    \left(O_m\right)_{ij} = \left(O_m^I\right)_{ij}+\left(O_m^{II}\right)_{ij},\label{p677}
\end{align}
where the individual terms are given by 
\begin{align}
    O_m^I = &\frac{i}{2}\begin{pmatrix}
    \mathbb 0 & \text{diag}\left(  d_\tau\omega g \right)\\ -\text{diag}\left( d_\tau\omega g \right) & \mathbb 0
    \end{pmatrix},\label{p678}\\
    O_m^{II} = & \frac{i}{2} \begin{pmatrix}
    d_\tau\omega & d_\tau\omega\\ d_\tau\omega& d_\tau\omega
    \end{pmatrix} \odot \begin{pmatrix}
-\left(\mathcal G^{[0]}\right)^{22}&\left(\mathcal G^{[0]}\right)^{21}\\ 
\left(\mathcal G^{[0]}\right)^{12}&-\left(\mathcal G^{[0]}\right)^{11}
\end{pmatrix}, \label{p679}
\end{align}
and we denote the Hadamard product between two equally-sized matrices $A$ and $B$ as $(A\odot B)_{ij}=A_{ij}B_{ij}$.
%--------------------------------------------------------------------
%--------------------------------------------------------------------
%--------------------------------------------------------------------
%--------------------------------------------------------------------
\section{Computing the tangent vectors \label{tangent}}
The explicit form of the tangent vectors $\ket{\Psi_P}=d_{\theta_P}\ket{\Psi_{\text{NGS}}}$ depends on whether $\theta_P$ describes a Gaussian parameter $\xi_{kl}$, or an non-Gaussian parameter $\omega_{kl}$. This leads to two different types of fermionic polynomials acting on a fermionic Gaussian state vector, which we present in Sections~\ref{derivative_gaussian} and \ref{derivative_non_gaussian}.
%-------------------------------------------------------
\subsection{Derivative of variational state with respect to Gaussian parameters:\label{derivative_gaussian}}
We consider the derivative of the variational state Ansatz with respect to a Gaussian parameter $\xi_{kl}$, which leads to
\begin{align}
d_{\xi_{kl}}\ket{\Psi_{\text{NGS}}} = \hat U_{\text{FA}}\left(d_{\xi_{kl}}\hat U_{\text{GS}}\right)\ket{0},\label{p446} 
\end{align}
since the derivative $d_{\xi_{kl}}$ commutes with the unitary $\hat U_{\text{FA}}$. We have
\begin{align}
d_{\xi_{kl}}\hat U_{\text{GS}} =&\frac{1}{4}\hat A^T\left(d_{\xi_{kl}}U_{m}\right)U_m^T\hat A\hat U_{\text{GS}},\label{p449}
\end{align}
where $\hat A=\left(\hat a_{1,1},\dots,\hat a_{1,N_f},\hat a_{2,1},\dots,\hat a_{2, N_f}\right)^T$, which follows from applying a similar strategy as in \cite{SHI2018245} where $d_\tau \hat U_{\text{GS}}$ was computed. Thus, Eq.~\eqref{p449} results in a quadratic polynomial of fermionic operators multiplied from the left to the Gaussian unitary $\hat U_{\text{GS}}$. One can evaluate $d_{\xi_{kl}}U_{m}= d_{\xi_{kl}}e^{i\xi}$ either using Wilcox's formula \cite{wilcox1967exponential}, or through diagonalization of the matrix $\xi$ \cite{tsai2003note}. We follow the latter approach here. Let $d_{\xi_1},\dots d_{\xi_{2N_f}}$ denote the eigenvalues of $\xi$, so that \cite{najfeld1995derivatives}
\begin{align}
e^{i\xi}=X_\xi\begin{pmatrix}
e^{id_{\xi_{1}}}&&\\
&\ddots&\\
&&e^{id_{\xi_{2N_f}}}
\end{pmatrix}X_\xi^{-1},\label{p450}
\end{align}
where $X_\xi$ is an $(2N_f\times 2N_f)$-dimensional matrix whose $j$-th column vector belongs to the $j$-th eigenvalue $d_{\xi_j}$. Since $\xi$ is a structured matrix (it is anti-symmetric and Hermitian), the matrix derivative has to preserve this symmetry in order to ensure that $d_{\xi_{kl}}U_m=-d_{\xi_{lk}}U_m$.  One can compute the derivative through \cite{tsai2003note}
\begin{align}
d_{\xi_{kl}}U_m =X_\xi V(\xi_{kl})X_\xi^{-1}\label{p447},
\end{align}
where $V(\xi_{kl})$ is a $(2N_f\times 2N_f)$-dimensional matrix whose $(m,n)$-th entry depends on the parameter $\xi_{kl}$ and is given by
\begin{align}
V(\xi_{kl})_{mn} = & \frac{1}{2}\left(X_{\xi}^{-1} (\ket{k}\bra{l}-\ket{l}\bra{k})X_\xi\right)_{mn}\begin{cases}
\frac{e^{id_{\xi_m}}-e^{id_{\xi_n}}}{d_{\xi_m}-d_{\xi_n}}&,\ \text{if} \ m\neq n,\\
ie^{id_{\xi_m}}&,\ \text{if} \ m=n.
\end{cases}\label{p448}
\end{align} 
Here, $\ket{k}$ ($\bra{l}$) is a column (row) vector of length $2N_f$ with a 1 at its $k$ ($l$)-th entry and zeros everywhere else. In order to arrive at Eq.~\eqref{p447}, we used that $\xi$ is skew-symmetric and thus $d_{\xi_{kl}}e^{i\xi}=-d_{\xi_{lk}}e^{i\xi}$. 

In the following, we bring Eq.~\eqref{p449} into a form that allows for the use of the analytical formula of Eq.~\eqref{p360}. We define the $(2N_f\times 2N_f)$-matrices
\begin{align}
k(\xi_{kl})=&\frac{1}{4}\left(d_{\xi_{kl}}U_{m}\right)U_m^T,\label{p598}\\
K(\xi_{kl}) =& \begin{pmatrix}
k(\xi_{kl})^{++}&k(\xi_{kl})^{+-}\\
k(\xi_{kl})^{-+}&k(\xi_{kl})^{--}
\end{pmatrix},
\end{align}
where the matrix elements of the block matrices appearing in the latter expression are defined in Eqs.~\eqref{p490}-\eqref{p492}. Using Wilcox's formula we can show that $k(\xi_{kl})$ defined in Eq.~\eqref{p598} is skew-symmetric, as we show in the following. By using the orthogonality property of $U_m=e^{i\xi}$, and $\left(d_{\xi_{kl}}\xi\right)^T = -d_{\xi_{kl}}\xi$, we compute \cite{wilcox1967exponential}
\begin{align}
    k(\xi_{kl})^T=&\frac{1}{4}e^{i\xi} \int du e^{-(1-u)i\xi}\left( d_{\xi_{kl}}\xi\right)^Te^{-ui\xi}=-\frac{1}{4}\int du e^{ui\xi}\left( d_{\xi_{kl}}\xi\right)e^{(1-u)i\xi}e^{-i\xi}=-k(\xi_{kl}),\label{p712}
\end{align}
and therefore $K(\xi_{kl})^T=-K(\xi_{kl})$. Using the transformation $\hat A=W_m \hat C$, where 
\begin{align}
W_m=&\begin{pmatrix}
\mathds 1_{N_f}&\mathds 1_{N_f}\\
-i\mathds 1_{N_f}&i\mathds 1_{N_f}
\end{pmatrix},\label{p654}
\end{align}
and the definitions from Eqs.~\eqref{p490}-\eqref{p492}, we can write Eq.~\eqref{p446} as 
\begin{align}
    d_{\xi_{kl}}\ket{\Psi_{\text{NGS}}} 
    =& -\hat U_{\text{FA}}\hat C^T K(\xi_{kl}) \hat C \ket{\Psi_{\text{GS}}},\label{p701}
\end{align}
where $\hat C=\left(\hat c_1,\dots,\hat c_{N_f},\hat c_1^\dag,\dots,\hat c_{N_f}^\dag\right)^T$. This equation allows for a straightforward application of Eq.~\eqref{p360} for the evaluation of the respective Gram matrix entries.
%-------------------------------------------------------
\subsection{Derivative of variational state with respect to non-Gaussian parameters:\label{derivative_non_gaussian}}
We now consider the derivative with respect to a non-Gaussian variational parameter $\omega_{kl}$, which leads to $d_{\omega_{kl}}\ket{\Psi_{\text{NGS}}} = \left(d_{\omega_{kl}}\hat U_{\text{FA}}\right)\ket{\Psi_{\text{GS}}}$. Using the definition of Eq.~\eqref{p9}, we get $d_{\omega_{kl}}\hat U_{\text{FA}}=\frac{i}{2}:\hat n_k\hat n_l:\hat U_{\text{FA}}=\frac{i}{2}\hat U_{\text{FA}}:\hat n_k\hat n_l:$. This is a quartic monomial of fermionic operators multiplied from the left (or right) to the non-Gaussian unitary $\hat U_{\text{FA}}$. 
%--------------------------------------------------------------------
%--------------------------------------------------------------------
%--------------------------------------------------------------------
%--------------------------------------------------------------------
\section{Derivations for the HITGD approach\label{math_deriv}}
We want to choose our Ansatz for the time evolution of the non-Gaussian variational parameters in such a way, that the second part of Eq.~\eqref{p556} vanishes, i.o.w.
\begin{align}
\frac{1}{8}\text{tr}\left(O_m^2\right)+\sum_{k,l=1}^{N_f}\frac{dE}{d\omega_{kl}}(d_\tau \omega)_{kl}\stackrel{!}{=}0.\label{p612}
\end{align}
Due to Eq.~\eqref{p562}, this is a sufficient condition in order for the energy to be monotonically decreasing. In order to sattisfy Eq.~\eqref{p612}, we have to evaluate the exact expression of the trace of the square of the mean-field non-Gaussian matrix given in Eq.~\eqref{p677} and solve for $d_\tau\omega_{kl}$. The square of the mean-field non-Gaussian matrix is given by
\begin{align}
\text{tr}\left(O_m^2\right)
=& \text{tr}\left(\left(O_m^{I}\right)^2\right)+2\text{tr}\left(O_m^{I}O_m^{II}\right)+\text{tr}\left(\left(O_m^{II}\right)^2\right),\label{p611}
\end{align}
where $O_m^I$ and $O_m^{II}$ are given by Eqs.~\eqref{p678} and \eqref{p679}, respectively. Using the symmetry of $\omega$, we have 
\begin{align}
    O_m^I = &\frac{i}{2}\begin{pmatrix}
    \mathbb 0 & \text{diag}\left(  d_\tau\omega g \right)\\ -\text{diag}\left( d_\tau\omega g \right) & \mathbb 0
    \end{pmatrix},\label{p1020}\\
    O_m^{II} = & \frac{i}{2} \begin{pmatrix}
    d_\tau\omega & d_\tau\omega\\ d_\tau\omega& d_\tau\omega
    \end{pmatrix} \odot \begin{pmatrix}
-\left(\mathcal G^{[0]}\right)^{22}&\left(\mathcal G^{[0]}\right)^{21}\\
\left(\mathcal G^{[0]}\right)^{12}&-\left(\mathcal G^{[0]}\right)^{11}
\end{pmatrix}. \label{p1021}
\end{align} 
We will compute the four terms appearing on the right-hand side of Eq.~\eqref{p611} individually. 
The trace over the first expression in Eq.~\eqref{p611} is given by 
\begin{align}
 \text{tr}\left(\left(O_m^I\right)^2\right)=\frac{1}{2}\sum_{k,l,m=1}^{N_f}(d_\tau\omega)_{kl}(gg^T)_{lm}(d_\tau\omega)_{mk}.\label{p1022}
\end{align}
We now turn to the cross terms and compute 
\begin{align}
O_m^IO_m^{II}=&\frac{i^2}{4}\begin{pmatrix}
\text{diag}\left(d_\tau\omega g\right)\left( d_\tau\omega\odot \left(\mathcal G^{[0]}\right)^{12} \right)&-\text{diag}\left(d_\tau\omega g\right)\left( d_\tau\omega\odot \left(\mathcal G^{[0]}\right)^{11} \right)\\
\text{diag}\left(d_\tau\omega g\right)\left( d_\tau\omega\odot \left(\mathcal G^{[0]}\right)^{22} \right)&-\text{diag}\left(d_\tau\omega g\right)\left( d_\tau\omega\odot \left(\mathcal G^{[0]}\right)^{21} \right)
\end{pmatrix}.\label{p614}
\end{align}
Since we are interested in computing the trace of Eq.~\eqref{p614}, we are  only considering the upper-left and lower-right blocks. Each of these blocks contains the matrix product of a diagonal matrix $\text{diag}\left(d_\tau\omega g \right)$ with the matrix that contains the Hadamard product $d_\tau\omega\odot \left(\mathcal G^{[0]}\right)^{12}$ and $-d_\tau\omega\odot \left(\mathcal G^{[0]}\right)^{21}$, respectively. The matrix product of some $(N\times N)$-diagonal matrix $\tilde D$ with an equally sized matrix $M$ is given by 
\begin{align}
\tilde DM=\begin{pmatrix}
\tilde D_{11}M_{11}& \tilde D_{11}M_{12}&\dots& \tilde D_{11}M_{1N}\\
\tilde D_{22}M_{21}& \tilde D_{22}M_{22}&\dots&\tilde D_{22}M_{2N}\\
\vdots&\vdots&\ddots&\vdots\\
\tilde D_{NN}M_{N1}& \tilde D_{NN}M_{N2}&\dots&\tilde D_{NN}M_{NN}
\end{pmatrix}.\label{p617}
\end{align}
In our case (i.e. Eq.~\eqref{p614}) the matrix $M$ corresponds to the Hadamard product $d_\tau\omega\odot \left(\mathcal G^{[0]}\right)^{12}$ in one instance and $-d_\tau\omega\odot \left(\mathcal G^{[0]}\right)^{21}$ in the other instance. Since the diagonal elements of $d_\tau\omega$ are all zero and due to the relation of Eq.~\eqref{p617}, we have $\text{tr}\left(O_m^IO_m^{II}\right)=0$. The third term is given by
\begin{align}
    \text{tr}\left(\left(O_m^{II}\right)^2\right)=\frac{1}{4}\sum_{k,l=1}^{N_f}\left((d_\tau \omega)_{k,l}\right)^2\left[\left(\left(\mathcal G^{[0]}\right)_{k,l}^{11}\right)^2+\left(\left(\mathcal G^{[0]}\right)_{k,l}^{12}\right)^2+\left(\left(\mathcal G^{[0]}\right)_{k,l}^{21}\right)^2+\left(\left(\mathcal G^{[0]}\right)_{k,l}^{22}\right)^2\right].\label{p683}
\end{align}
We define the four-dimensional tensor $\mathcal B$ with entries
\begin{align}
    \mathcal B_{klmn} =& \frac{1}{8}\left[(gg^T)_{lm}\delta_{nk}+(gg^T)_{ln}\delta_{mk}+(gg^T)_{km}\delta_{nl}+(gg^T)_{kn}\delta_{ml}\right]\nonumber\\&+\frac{1}{8}\left[\left(\left(\mathcal G^{[0]}\right)_{k,l}^{11}\right)^2+\left(\left(\mathcal G^{[0]}\right)_{k,l}^{12}\right)^2+\left(\left(\mathcal G^{[0]}\right)_{k,l}^{21}\right)^2+\left(\left(\mathcal G^{[0]}\right)_{k,l}^{22}\right)^2\right]\left(\delta_{mk}\delta_{nl}+\delta_{nk}\delta_{ml}\right),\label{p1030}
\end{align}
which allows us to write 
\begin{align}
    \text{tr}\left(O_m^2\right) = \sum_{k,l,m,n=1}^{N_f}(d_\tau\omega)_{kl}\mathcal B_{klmn}(d_\tau\omega)_{mn}.\label{p1031}
\end{align}
By definition, we have the symmetries $\mathcal B_{klmn}=\mathcal B_{lkmn}=\mathcal B_{klnm}=\mathcal B_{lknm}$ and $\mathcal B_{klmn}=\mathcal B_{mnkl}$. We set $\mathcal B_{kklm}=0=\mathcal B_{klmm}$, since $(d_\tau\omega)_{kk}=0$ throughout the evolution. Then, Eq.~\eqref{p612} can be written as 
\begin{align}
    \frac{1}{8}\sum_{k,l,m,n=1}^{N_f}(d_\tau \omega)_{kl}\mathcal B_{klmn}(d_\tau \omega)_{mn}
    +\sum_{k,l=1}^{N_f}(d_\tau \omega)_{kl}\frac{dE}{d\omega_{kl}}\stackrel{!}{=}0.
\end{align}
By choosing
\begin{align}
    \frac{1}{8}\sum_{m,n=1}^{N_f}\mathcal B_{klmn}(d_\tau \omega)_{mn}=-
    \frac{dE}{d\omega_{kl}},\label{p1033}
\end{align}
for all $j,k\in\{1,\dots,n_f\}$, we are guaranteed that the left-hand side of Eq.~\eqref{p612} is identical to zero.  We reshape the $(N_f\times N_f\times N_f\times N_f)$-tensor $\mathcal B_{jklm}$ into a $(N_f^2\times N_f^2)$-matrix $\mathcal B_{(jk),(lm)}$ with tuple entries $(jk)$ and $(lm)$. Similarly, we reshape the $(N_f\times N_f)$-matrix $dE/d\omega$ into a column vector $E_{(jk)}$ of length $N_f^2$. In the following, we denote the tuple indices with capital roman letters, as in the reshaped representation of Eq.~\eqref{p410}. Then, Eq.~\eqref{p1033} reads
\begin{align}
    \frac{1}{8}\sum_{L=1}^{N_f^2}\mathcal B_{JL}(d_\tau\omega)_L = -\frac{dE}{d\omega_J}.
\end{align}
This equation can be solved for $d_\tau\omega$ by multiplying both sides of the equation with the Moore-Penrose inverse $\mathcal B^+$ of the reshaped tensor $\mathcal B$. This leads to
\begin{align}
    (d_\tau\omega)_K = -8\sum_{L=1}^{N_f^2}\mathcal B^+_{KL}\frac{dE}{d\omega_{L}}.\label{p1035}
\end{align}
By choosing the variational parameters according to Eq.~\eqref{p1035}, we are guaranteed a monotonically decreasing energy during the HITGD evolution. 
%--------------------------------------------------------------------
%--------------------------------------------------------------------
%--------------------------------------------------------------------
%--------------------------------------------------------------------
\section{Derivations for the energy gradient\label{en_grad}}
In this appendix, we derive the terms that lead to Eq.~\eqref{p649}. We will compute the resulting expressions seperately for the one-, and two-body Hamiltonian terms in Sections~\ref{contributions_1} and \ref{contributions_2}.
%-------------------------------------------------------
\subsection{Contributions from rotated one-body Hamiltonian terms\label{contributions_1}}
In order to compute the commutator $[\hat H_{\text{FA}},\hat c_i^\dag \hat c_j^\dag \hat c_i\hat c_j]$ in Eq.~\eqref{p648}, we first turn to terms that belong to the rotated one-particle Hamiltonian,
\begin{align}
\left[e^{i\sum_k\alpha_{pq}^{\text{FA}}(k)\hat n_{k}}\hat c_p^\dag \hat c_q, \hat c_i^\dag \hat c_j^\dag \hat c_i\hat c_j\right]=&e^{i\sum_k\alpha_{pq}^{\text{FA}}(k)\hat n_{k}}\left(-\hat c_i^\dag \hat c_j^\dag \hat c_i\hat c_q\delta_{p,j}+\hat c_i^\dag \hat c_j^\dag \hat c_j\hat c_q\delta_{p,i}-\hat c_p^\dag \hat c_i^\dag \hat c_i\hat c_j\delta_{j,q}+\hat c_p^\dag\hat  c_j^\dag \hat c_i\hat c_j\delta_{i,q}\right).\label{p465}
\end{align} 
In order to get the contributions due to the one-body terms, we will have to include the appropriate coefficients and sum over the fermionic modes, resulting in
\begin{align}
\sum_{p,q=1}^{N_f}f_{pq}^{\text{FA}}\left[e^{i\sum_k\alpha_{pq}^{\text{FA}}(k)\hat n_{k}} \hat c_p^\dag \hat c_q, \hat c_i^\dag \hat c_j^\dag \hat c_i\hat c_j\right]
%%%%%%%%%%%%%%%%%%%%%%%%
%=&-\sum_{p=1}^{N_f}f_{jp}^{\text{FA}}e^{i\sum_k\alpha_{jp}^{\text{FA}}(k) \hat n_{k}}\hat c_i^\dag \hat c_j^\dag \hat c_i\hat c_p+\sum_{p=1}^{N_f}f_{ip}^{\text{FA}}e^{i\sum_k\alpha_{ip}^{\text{FA}}(k) \hat n_{k}}\hat c_i^\dag \hat c_j^\dag \hat c_j\hat c_p\nonumber\\&-\sum_{p=1}^{N_f}f_{pj}^{\text{FA}}e^{i\sum_k\alpha_{pj}^{\text{FA}}(k) \hat n_{k}}\hat c_p^\dag \hat c_i^\dag \hat c_i\hat c_j +\sum_{p=1}^{N_f}f_{pi}^{\text{FA}}e^{i\sum_k\alpha_{pi}^{\text{FA}}(k) \hat n_{k}}\hat c_p^\dag \hat c_j^\dag \hat c_i\hat c_j\nonumber\\
%%%%%%%%%%%%%%%%%%%%%%%%
=&\left(\sum_{p=1}^{N_f}f_{ip}^{\text{FA}}e^{i\sum_k\alpha_{ip}^{\text{FA}}(k)\hat n_{k}}\hat c_i^\dag \hat c_j^\dag \hat c_j\hat c_p+(i\leftrightarrow j)\right)\nonumber\\&+\left(\sum_{p=1}^{N_f}f_{pi}^{\text{FA}}e^{i\sum_k\alpha_{pi}^{\text{FA}}(k)\hat n_{k}}\hat c_p^\dag \hat c_j^\dag \hat c_i\hat c_j+(i\leftrightarrow j)\right).\label{p467}
\end{align}
Using the following identity,
\begin{align}
\left(f_{ip}^{\text{FA}}\left\langle \Psi_{\text{GS}}\right|e^{i\sum_k\alpha_{ip}^{\text{FA}}(k) \hat n_{k}}\hat c_i^\dag \hat c_j^\dag \hat c_j\hat c_p\left|\Psi_{\text{GS}}\right\rangle\right)^*
%%%
=&-f_{pi}^{\text{FA}}\left\langle\Psi_{\text{GS}}\right|  e^{i\sum_k\alpha_{pi}^{\text{FA}}(k) \hat n_{k}}\hat c_p^\dag \hat c_j^\dag \hat c_i \hat c_j \left|\Psi_{\text{GS}}\right\rangle,\label{p572}
\end{align} 
the expectation value with respect to the fermionic Gaussian state of the terms in Eq.~\eqref{p467} may be written as 
\begin{align}
\left\langle\Psi_{\text{GS}}\right| \left[\hat H_{\text{FA}}^{(1)},\hat c_i^\dag \hat c_j^\dag \hat c_i\hat c_j\right]\left|\Psi_{\text{GS}} \right\rangle
%%%%%%%%%%%%%%%%%%%%%%%%
=&\left(\sum_{p=1}^{N_f}f_{ip}^{\text{FA}}\left\langle\Psi_{\text{GS}}\right| e^{i\sum_k\alpha_{ip}^{\text{FA}}(k)\hat  n_{k}}\hat c_i^\dag \hat c_j^\dag \hat c_j\hat c_p\left|\Psi_{\text{GS}}\right\rangle \ - \text{H.c.}\right)  +(i\leftrightarrow j), \label{p573}
\end{align}
where H.c. denotes the Hermitian conjugated of the first expression inside the big brackets on the right-hand side of Eq.~\eqref{p573}. The notation $(i\leftrightarrow j)$ is a placeholder for a duplicate of the expression inside the large brackets, but with the role of index $i$ replaced by $j$ (and vice versa). The result of Eq.~\eqref{p573} gives a purely imaginary number.
\subsection{Contributions from rotated two-body Hamiltonian terms\label{contributions_2}}
In this subsection, we will compute 
\begin{align}
\left\langle\Psi_{\text{GS}}\right| \left[\hat H_{\text{FA}}^{(2)},\hat c_i^\dag \hat c_j^\dag \hat c_i\hat c_j\right]\left|\Psi_{\text{GS}} \right\rangle=&
\frac{1}{2}\sum_{p,q,r,s=1}^{N_f}h_{pqrs}^{\text{FA}}\left\langle\Psi_{\text{GS}}\right| \left[e^{i\sum_k\beta_{pqrs}^{\text{FA}}(k)\hat  n_{k}} \hat  c_p^\dag \hat c_q^\dag \hat c_r \hat c_s,\hat c_i^\dag \hat c_j^\dag \hat c_i\hat c_j\right]\left|\Psi_{\text{GS}}\right\rangle.\label{p579}
\end{align}
Similarly to the previous subsection, one can use the identities
\begin{align}
\left(\sum_{p,q,r=1}^{N_f}h_{ipqr}^{\text{FA}}\left\langle\Psi_{\text{GS}}\right| e^{i\sum_k\beta_{ipqr}^{\text{FA}}(k) \hat n_{k}}\hat c_j^\dag \hat c_i^\dag \hat c_p^\dag \hat c_j \hat c_q\hat c_r\left|\Psi_{\text{GS}}\right\rangle\right)^*
%%%
=&-\sum_{p,q,r=1}^{N_f}h_{pqri}^{\text{FA}}\left\langle\Psi_{\text{GS}}\right| e^{i\sum_k\beta_{pqri}^{\text{FA}}(k) \hat n_{k}}\hat c_p^\dag \hat c_q^\dag \hat c_j^\dag \hat c_r \hat c_j\hat c_i\left|\Psi_{\text{GS}}\right\rangle,\label{p576}\\
%%%
\left(\frac{1}{2}\sum_{p,q=1}^{N_f}h_{ijpq}^{\text{FA}}\left\langle\Psi_{\text{GS}}\right| e^{i\sum_k\beta_{ijpq}^{\text{FA}}(k) \hat n_{k}}\hat c_i^\dag \hat c_j^\dag  \hat c_p\hat c_q\left|\Psi_{\text{GS}}\right\rangle\right)^*=&
-\frac{1}{2}\sum_{p,q=1}^{N_f}h_{pqij}^{\text{FA}}\left\langle\Psi_{\text{GS}}\right| e^{i\sum_k\beta_{pqij}^{\text{FA}}(k)\hat  n_{k}}\hat c_p^\dag \hat c_q^\dag  \hat c_j\hat c_i\left|\Psi_{\text{GS}}\right\rangle.\label{p577}
\end{align}
to arrive at the following expression,
\begin{align}
\left\langle \Psi_{\text{GS}}\right|\left[\hat H_{\text{FA}}^{(2)},\hat c_i^\dag \hat c_j^\dag \hat c_i\hat c_j\right]\left|\Psi_{\text{GS}} \right\rangle
%%%
=&\left[\left(\sum_{p,q,r=1}^{N_f}h_{ipqr}^{\text{FA}}\left\langle\Psi_{\text{GS}}\right| e^{i\sum_k\beta_{ipqr}^{\text{FA}}(k) \hat n_{k}}\hat c_j^\dag \hat c_i^\dag \hat c_p^\dag \hat c_j \hat c_q\hat c_r\left|\Psi_{\text{GS}}\right\rangle-\ \text{H.c.}\right)\right.\nonumber\\
&\left.+\frac{1}{2}\left(\sum_{p,q=1}^{N_f}h_{ijpq}^{\text{FA}}\left\langle\Psi_{\text{GS}}\right| e^{i\sum_k\beta_{ijpq}^{\text{FA}}(k) \hat n_{k}}\hat c_i^\dag \hat c_j^\dag  \hat c_p\hat c_q\left|\Psi_{\text{GS}}\right\rangle-\ \text{H.c.}\right)\right]+(i\leftrightarrow j).\label{p578}
\end{align}
Note, that the resulting number in Eq.~\eqref{p578} is purely imaginary. We can simplify part of the above expression through
\begin{align}
\frac{1}{2}\sum_{p,q=1}^{N_f}h_{ijpq}^{\text{FA}}\left\langle\Psi_{\text{GS}}\right| e^{i\sum_k\beta_{ijpq}^{\text{FA}}(k) \hat n_{k}}\hat c_i^\dag \hat c_j^\dag  \hat c_p\hat c_q\left|\Psi_{\text{GS}}\right\rangle+(i\leftrightarrow j)=2\sum_{p<q}^{N_f}h_{ijpq}^{\text{FA}}\left\langle\Psi_{\text{GS}}\right| e^{i\sum_k\beta_{ijpq}^{\text{FA}}(k) \hat n_{k}}\hat c_i^\dag \hat c_j^\dag  \hat c_p\hat c_q\left|\Psi_{\text{GS}}\right\rangle.\label{p594}
\end{align}
The results from Subsections \ref{contributions_1}-\ref{contributions_2} lead to Eq.~\eqref{p649}.
%--------------------------------------------------------------------
%--------------------------------------------------------------------
%--------------------------------------------------------------------
%--------------------------------------------------------------------
\end{document}